\documentclass[]{imsart}

\RequirePackage[numbers]{natbib}
\RequirePackage{amsthm,amsmath}
\usepackage{pictexwd}
\usepackage{amssymb,graphicx,url}
\usepackage[colorlinks,citecolor=blue,urlcolor=blue]{hyperref}
\usepackage{bm}

\startlocaldefs    
\def\a{\alpha}
\def\b{\beta}
\def\R{\mathbb R}

\def\d{\delta}

\def\E{{\mathbb E}}

\def\P{{\mathbb P}}

\def\l{\lambda}

\def\labda1{\lambda_1}
\def\labda2{\lambda_2}
\def\bma{\bm\alpha}
\def\bmb{\bm\beta}
\def\m{\mu}

\def\e{\varepsilon}
\def\f{\phi}
\def\t{\tau}

\def\comment#1{\relax}

\def\=in{\mathop{\rm =}}

\newtheorem{theorem}{Theorem}[section]

\newtheorem{lemma}{Lemma}[section]

\newtheorem{remark}{Remark}[section]

\numberwithin{equation}{section}
\theoremstyle{plain}
\setattribute{journal}{name}{}
\endlocaldefs

\usepackage{amsmath,here,amsfonts,latexsym,sym1,graphicx,here}
\usepackage{amsthm,curves}
\usepackage{graphicx}
\usepackage{caption}
\usepackage{subcaption}
\usepackage{float}

\advance \topmargin by -\headheight \advance \topmargin by
-\headsep

\def\a{\gamma}
\def\m{\mu}
\def\bmb{\bm \b}
\def\bmB{\bm B}
\def\bmI{\bm I}

\def\GG{\Gamma}

\def\P{{\mathbb P}}

\def\bmX{\bm X}
\def\bmx{\bm x}
\def\bmv{\bm v}

\def\a{\alpha}

\begin{document}
\begin{frontmatter}
\title{The Lagrange approach in the monotone single index model}
\runtitle{Single index model}

\begin{aug}
\runauthor{Piet Groeneboom}
\author{\fnms{Piet} \snm{Groeneboom}\corref{}\ead[label=e1]{P.Groeneboom@tudelft.nl}
\ead[label=u1,url]{http://dutiosc.twi.tudelft.nl/\textasciitilde pietg/}}
\address{Delft University of Technology, Building 28, Van Mourik Broekmanweg 6, 2628 XE Delft, The Netherlands.\\ \printead{e1}}
\end{aug}

\begin{abstract}
The finite-dimensional parameters of the monotone single index model are often estimated by minimization of a least squares criterion and reparametrization to deal with the non-unicity. We avoid the reparametrization by using a Lagrange-type method and replace the minimization over the finite-dimensional parameter $\bma$ by a ``crossing of zero'' criterion at the derivative level. In particular, we consider a simple score estimator (SSE), an efficient score estimator (ESE), and a penalized least squares estimator (PLSE) for which we can apply this method. The SSE and ESE were discussed in \cite{balabdaoui2018}, but the proofs still used reparametrization. Another version of the PLSE was discussed in \cite{Kuchibhotla_patra:17}, where also reparametrization was used. The estimators are compared with  the profile least squares estimator (LSE), Han's maximum rank estimator (MRE), the effective dimension reduction estimator (EDR) and a linear least squares estimator, which can be used if the covariates have an elliptically symmetric distribution. We also investigate the effects of random starting values in the search algorithms.
\end{abstract}

\begin{keyword}[class=AMS]
\kwd[Primary ]{62G05}
\kwd{62N01}
\kwd[; secondary ]{62-04}
\end{keyword}

\begin{keyword}
\kwd{single index model, natural cubic smoothing splines, LSE, Lagrange}
\end{keyword}

\end{frontmatter}

\section{Introduction}
\label{sec:intro}
The monotone single index model tries to predict a response from the linear combination of a finite number of parameters and a function linking this linear combination of parameters to the response via a monotone {\it link function} $\psi_0$ which is  unknown. So, more formally, we have the model
\begin{align*}
Y=\psi_0(\bma_0^T\bmX)+\e,
\end{align*}
where $Y $ is a one-dimensional random variable, $\bm X = (X_1,\ldots, X_d)^T$ is a $d$-dimensional random vector with distribution function $G$,  $\psi_0$ is monotone  and $\e$ is a one-dimensional random variable such that $\E[\e | \bm X] = 0$ $G$-almost surely. The regression parameter $\bma_0$ is a vector  of norm $\|\bma_0\|_2=1$, where $\|\,  \cdot \|_2$ denotes the Euclidean norm in $\R^d$.

The profile least squares estimate of $\bma_0$ is an $M$ estimate in two senses: for fixed $\bma$ the least squares criterion
\begin{align}
\label{step1}
\psi\mapsto n^{-1}\sum_{i=1}^n\left\{Y_i-\psi(\bma^T\bmX_i)\right\}^2
\end{align}
is minimized for all monotone functions $\psi$ (either decreasing or increasing) which gives an $\bma$ dependent function $\hat\psi_{n,\bma}$, and the function
\begin{align}
\label{step2}
\bma\mapsto n^{-1}\sum_{i=1}^n\left\{Y_i-\hat\psi_{n,\bma}(\bma^T\bmX_i)\right\}^2
\end{align}
is minimized over $\bma$. This gives the profile least squares estimator $\hat\bma_n$ of $\bma_0$, which we will call LSE in the sequel.
Although this estimate of $\bma_0$ has been  known now for a very long time (more than 30 years probably), it is not known whether it is $\sqrt{n}$ convergent (under appropriate regularity conditions), let alone that its asymptotic distribution is known. Also, simulation studies are rather inconclusive. For example, it is conjectured in \cite{tanaka2008} on the basis of simulations that the rate of convergence of $\hat\bma_n$ is $n^{9/20}$. Other simulation studies were presented in \cite{balabdaoui2018} which suggest that even if the LSE is $\sqrt{n}$ convergent under some conditions (for example elliptic symmetry of the distribution of the covariate $\bmX$), there are other estimates which have a better performance in these circumstances.

In order to get more insight in the matter, a combination of an $M$ estimate and an $L$ estimate was studied in \cite{balabdaoui2018}.
For fixed $\bma$ we use the least squares estimator $\hat \psi_{n,\bma}$ of the link function, minimizing (\ref{step1}) (the $M$ part), just as is done for the LSE, but we try to find the estimate of $\bma_0$ by solving an equation in $\bma$ for these least squares functions $\hat \psi_{n,\bma}$ (the $L$ part) instead of minimizing again over $\bma$.
This  means, using the Lagrange approach for the side condition that $\|\bma\|_2=1$, that in principle we want to solve an equation of the type
\begin{align}
\label{L_criterion}
\frac1n\sum_{i=1}^n \bigl\{\hat\psi_{n,\bma}(\bma^T\bmX)-Y_i\bigr\}\left\{\bmX_i\frac{d}{du}\hat\psi_{n,\bma}(u)\bigr|_{u=\bma^T\bmX_i}  +\frac{\partial}{\partial\bma}\hat\psi_{n,\bma}(u)\bigr|_{u=\bma^T\bmX_i}\right\}+\l\bma=\bm0,
\end{align}
(differentiating both w.r.t.\ the argument of the function $x\mapsto\hat\psi_{n,\bma}(\bma^T\bmx)$ and w.r.t. its parameter $\bma$), where the Lagrange multiplier $\l$ satisfies
\begin{align}
\label{Lagrange_equation}
\l=-\frac1n\sum_{i=1}^n \bigl\{\hat\psi_{n,\bma}(\bma^T\bmX_i)-Y_i\bigr\}\bma^T\left\{\bmX_i\frac{d}{du}\hat\psi_{n,\bma}(u)\bigr|_{u=\bma^T\bmX_i}  +\frac{\partial}{\partial\bma}\hat\psi_{n,\bma}(u)\bigr|_{u=\bma^T\bmX_i}\right\},
\end{align}
using the condition $\bma^T\bma=1$. Note that the left-hand side of (\ref{L_criterion}) is $1/2$ times the derivative w.r.t.\ $\bma$ of the least squares criterion
\begin{align}
\label{M-criterion}
\frac1n\sum_{i=1}^n \bigl\{\hat\psi_{n,\bma}(\bma^T\bmX)-Y_i\bigr\}^2+\l\left\{\|\bma\|^2_2-1\right\},
\end{align}
where one would try to choose the Lagrange multiplier $\l$ in such a way that $\|\bma\|_2^2=1$, if $\bma$ is the minimizer of (\ref{M-criterion}).

Eliminating $\l$ using (\ref{Lagrange_equation}), this would give the following equation in $\bma$:
\begin{align}
\label{fundamental_eq}
\frac1n\left(\bm I-\bma\bma^T\right)\sum_{i=1}^n \bigl\{\hat\psi_{n,\bma}(\bma^T\bmX)-Y_i\bigr\}\left\{\bmX_i\frac{d}{du}\hat\psi_{n,\bma}(u)\bigr|_{u=\bma^T\bmX_i}  +\frac{\partial}{\partial\bma}\hat\psi_{n,\bma}(u)\bigr|_{u=\bma^T\bmX_i}\right\}=\bm0.
\end{align}
The derivatives in the last factor of the left-hand side of this expression do not exist for the direct least squares estimate $\hat\psi_{n,\bma}$. But the equation corresponds in the underlying model to the equation
\begin{align}
\label{fundamental_eq2}
&\left(\bm I-\bma\bma^T\right)\E\left[\bigl\{\psi_{\bma}(\bma^T\bmX)-Y\bigr\}\left\{\bmX\frac{d}{du}\psi_{\bma}(u)\bigr|_{u=\bma^T\bmX}  +\frac{\partial}{\partial\bma}\psi_{\bma}(u)\bigr|_{u=\bma^T\bmX}\right\}\right]\nonumber\\
&=\left(\bm I-\bma\bma^T\right)\E \bigl\{\psi_0(\bma_0^T\bmX)|\bma^TX)-\psi_0(\bma_0^T\bmX)\bigr\}\bmX\frac{d}{du}\psi_{\bma}(u)\bigr|_{u=\bma^T\bmX}
=\bm0,
\end{align}
where $u\mapsto\psi_{\bma}(u)=E\{\psi_0(\bma_0^T\bmx)|\bma^T\bmx=u)$ minimizes
\begin{align}
\label{model_LS}
\psi\mapsto\int\left\{\psi_0(\bma_0^T\bmx)-\psi(\bma^T\bmx)\right\}^2\,dG(\bmx),
\end{align}
over nondecreasing functions $\psi$ on $\R$ (see Proposition 3 in \cite{balabdaoui2018}), if $G$ is the distribution function of the covariate $\bmX$, where we can assume that the derivatives in (\ref{fundamental_eq2}) exist.
The first equality in (\ref{fundamental_eq2}) follows from
\begin{align*}
&\E \bigl\{\psi_0(\bma_0^T\bmX)|\bma^TX)-\psi_0(\bma_0^T\bmX)\bigr\}\frac{\partial}{\partial\bma}\psi_{\bma}(u)\bigr|_{u=\bma^T\bmX}\\
&=\E\left[\frac{\partial}{\partial\bma}\psi_{\bma}(u)\bigr|_{u=\bma^T\bmX}\E\left(\left\{\psi_0(\bma_0^T\bmX)|\bma^TX)-\psi_0(\bma_0^T\bmX)\right\}\Bigm|\bma^TX\right)\right]\\
&=\bm0.
\end{align*}
and expresses that $\psi_{\bma}$ is the minimizer of (\ref{model_LS}) in the model, for fixed $\bma$.

One can show that the sample equivalent of equation (\ref{fundamental_eq2}), but without the derivative, yields a consistent and $\sqrt{n}$ convergent estimate of the regression parameter $\bma_0$. So the {\it Simple Score Estimator} (SSE) is defined (approximately) as the value of $\bma$ such that
\begin{align}
\label{SSE_eq}
\frac1n\left(\bm I-\bma\bma^T\right)\sum_{i=1}^n \bigl\{\hat\psi_{n,\bma}(\bma^T\bmX_i)-Y_i\bigr\}\bmX_i=\bm0,
\end{align}
where we add ``approximately'', since we cannot hope to achieve exact equality to zero because of the discrete character of the left-hand side of the equation. Instead we look for a ``crossing of zero'' of the left-hand side, as explained in \cite{balabdaoui2018}. In practice, this means that we minimize the norm of the left-hand side as a function of $\bma$ in the computer program for computing the SSE, see \cite{github:18}. Note that we do not need any tuning parameter for this estimator.

The method of computing $\hat\bma_n$ via (\ref{SSE_eq}) was introduced in \cite{balabdaoui2018}, but  consistency and asymptotic normality of the estimate defined in this way were not proved. Instead, the proofs of these facts were based on a lower dimensional parametrization and computation of the estimate via the lower dimensional parametrization,  involving the Jacobian of the transformation from $\R^{d-1}$ to $\R^d$, and not using the Lagrange approach. But in fact, many of the computations in the simulations were done using the Lagrange approach after we noticed empirically that these gave the same results in our simulations.

It is also possible to use the least squares estimator $\hat\psi_{n,\bma}$ in the sample equivalent of  (\ref{fundamental_eq2}), and use an estimate of the derivative $\frac{d}{du}\psi_{\bma}(u)\bigr|_{u=\bma^T\bmX}$, constructed from $\hat\psi_{\bma}$. This estimate is defined by:
\begin{align}
\label{estimate_derivative_psi}
\tilde\psi_{n,h,\bm\a}'(u)=\frac1h\int K\left(\frac{u-x}h\right)\,d \hat\psi_{n,\bm\a}(x),
\end{align}
where $K$ is one of the usual kernels, symmetric around zero and with support $[-1,1]$, and where $h$ is a bandwidth of order $n^{-1/7}$ for sample size $n$. Note that this estimate is rather different from the derivative of a Nadaraya-Watson estimate which is also used in this context and which is in fact the derivative of a ratio of two kernel estimates.
Replacing (\ref{SSE_eq}) by the equation
\begin{align}
\label{ESE_eq}
\frac1n\left(\bm I-\bma\bma^T\right)\sum_{i=1}^n \bigl\{\hat\psi_{n,\bma}(\bma^T\bmX_i)-Y_i\bigr\}\bmX_i\tilde\psi_{n,h,\bm\a}'(u)\bigr|_{u=\bma^T\bmX_i}=\bm0,
\end{align}
then yields, under the appropriate regularity conditions, the so-called {\it efficient score estimator} (ESE), which is not only consistent and square root $n$ consistent, but also asymptotically efficient (see \cite{balabdaoui2018}).

So, surprisingly, the replacement of the  minimization (\ref{step2}) in the profile least squares estimator by the search of the solution of  (\ref{SSE_eq}) or  (\ref{ESE_eq}) in $\bma$ produces a $\sqrt{n}$ convergent estimator $\hat\bma_n$, which is even efficient if we use (\ref{ESE_eq}).

We can also use as our estimator of the link function a cubic spline $\hat\psi_{n,\bma}$, which is defined as the function minimizing
\begin{align}
\label{spline}
\sum_{i=1}^n\left\{f(\bma^T\bmX_i)-Y_i\right\}^2+\mu\int_a^b f''(x)^2\,dx,
\end{align}
over the class of functions ${\cal S}_2[a,b]$ of differentiable functions $f$ with an absolutely continuous first derivative, where
\begin{align*}
a=\min_i\bma^T\bmX_i,\qquad b=\max_i\bma^T\bmX_i,
\end{align*}
see \cite{green_silverman:94}, pp. 18 and 19, where $\m>0$ is the penalty parameter. Using these estimators of the link function, the estimate $\hat\bma_n$ of $\bma_0$ is then found in \cite{Kuchibhotla_patra:17} by using a $(d-1)$-dimensional parametrization $\b$ and a transformation $S:\bm\beta\mapsto S(\bmb)=\bma$, where $S(\bmb)$ belongs to the surface of the unit sphere in $\R^d$, and minimizing the criterion
\begin{align*}
\bmb\mapsto\sum_{i=1}^n\{Y_i-\hat\psi_{S(\bmb),\lambda}(S(\bmb)^T\bmX_i)\}^2,
\end{align*}
over $\bmb$, where $\hat\psi_{S(\bmb),\lambda}$ minimizes (\ref{spline}) for fixed $\bma=S(\bmb)$.

Analogously to our approach above we can skip the reparametrizion, and, moreover, can try to solve directly the equation
\begin{align}
\label{spline_eq}
\frac1n\left(\bm I-\bma\bma^T\right)\sum_{i=1}^n \bigl\{\hat\psi_{n,\bma,\l}(\bma^T\bmX_i)-Y_i\bigr\}\bmX_i\tilde\psi_{n,\bm\a,\l}'(u)\bigr|_{u=\bma^T\bmX_i}=\bm0,
\end{align}
where $\tilde\psi_{n,\bma,\lambda}$ minimizes (\ref{spline}) for fixed $\bma$ and $\tilde\psi'_{n,\bma,\lambda}$ is its derivative. The  method can also be used for not necessarily monotone functions $\psi_0$. We call this estimator the Penalized Least Squares Estimator (PLSE).

Another important point not addressed in \cite{balabdaoui2018} is the starting value of the iterations in the computer program. It is rather essential to have a starting point for the search of the regression parameter which is sufficiently close to the real underlying parameter $\bma_0$, in particular for the efficient score estimator and spline estimator. The simulations in \cite{balabdaoui2018} were therefore done by starting at $\bma_0$, something one can of course not do in practice. We therefore investigate how the methods perform if one uses random starting values. To this end we first compute the (cube root $n$ consistent, see \cite{Balabdaoui_Durot_Jankowski:2016}) profile least squares estimate of $\bma_0$, and next use this as a starting value for the algorithms for the SSE, ESE and PLSE.

It is the purpose of the present paper to provide the changes in the proofs of the consistency and asymptotic distribution of the SSE, ESE and PLSE, computed via the Lagrange-type method given above, but without going through all technicalities again, which are given in \cite{balabdaoui2018} and \cite{Kuchibhotla_patra:17}. So we will give the essential changes in the proofs and refer for the needed entropy machinery to these papers. For the simple score estimator (SSE) we go through all the needed steps, though, as an example of the skeleton of the proof for the other estimators. We further illustrate the validity of the Lagrange approach by simulations and also show that one can start the algorithms at a random starting point.

\section{An illustrative example}
\label{sec:example}

\begin{figure}[!ht]
	\centering
	\begin{subfigure}{0.45\linewidth}
		\includegraphics[width=0.95\textwidth]{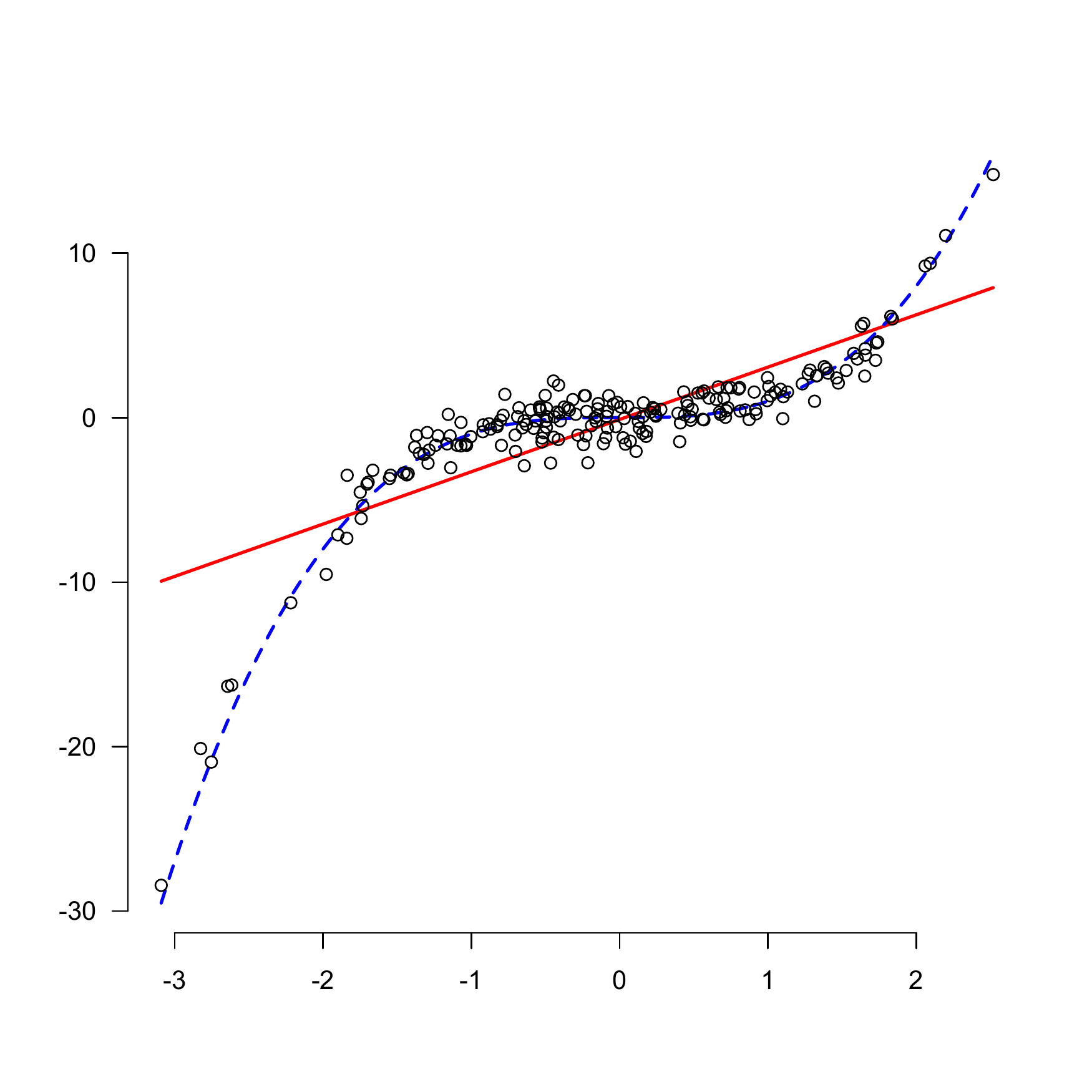}
		\caption{linear estimate}
	\end{subfigure}
	\begin{subfigure}{0.45\linewidth}
	\includegraphics[width=0.95\textwidth]{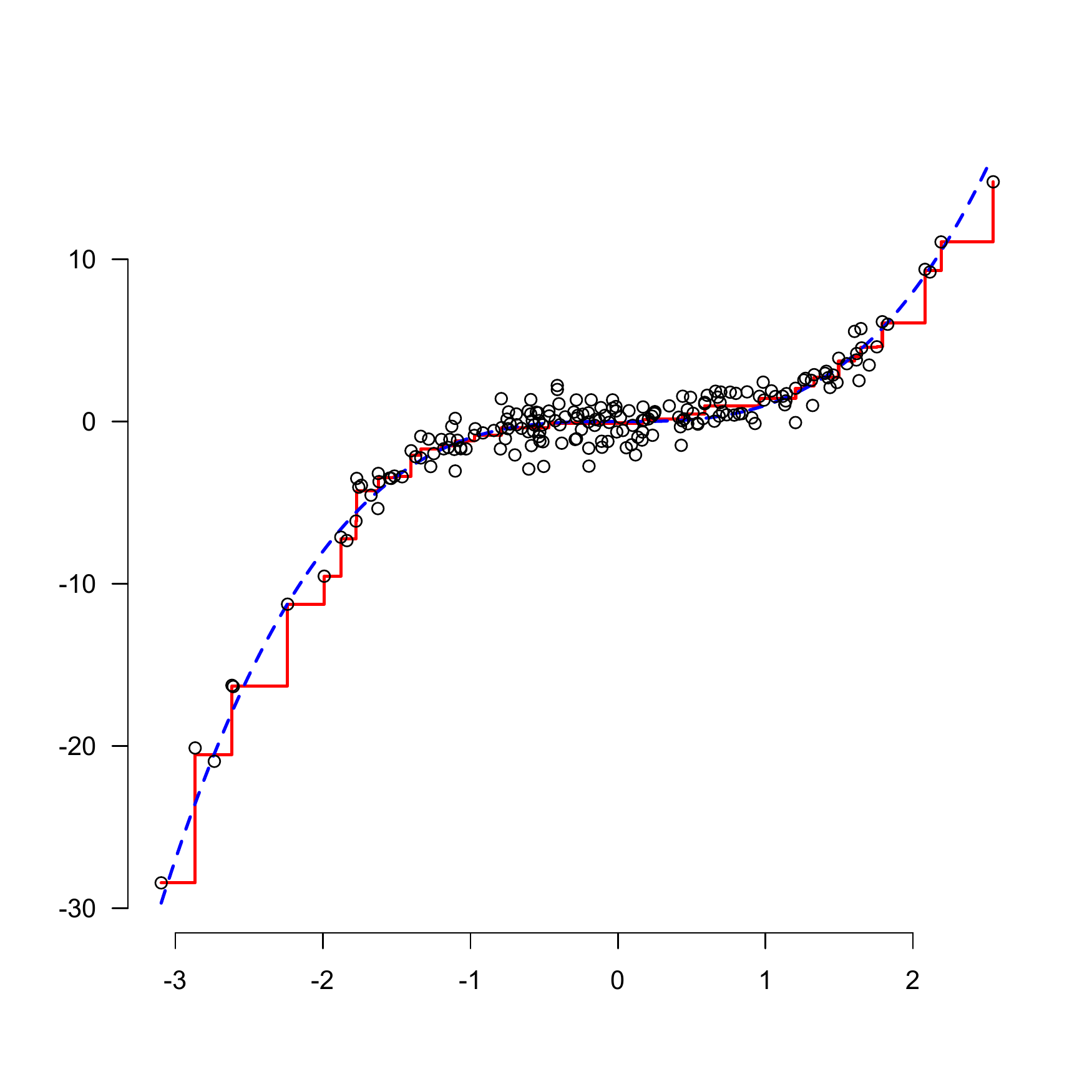}
	\caption{profile least squares estimate}
	\end{subfigure}
	\caption{A linear and non-linear estimate of the link function $\psi_0(x)=x^3$ for the model  (\ref{first_model}), where the sample size $n=200$. In (a) the estimate is given by the line (solid) and in $(b)$ by the step function (solid), which is the profile least squares estimate. The dashed curve is the function $\psi_0$.}
	\label{fig:linear_fit_comparison}
\end{figure}

We consider the following model, that was also studied in \cite{balabdaoui2018}:
\begin{align}
\label{first_model}
Y = \psi_0(\bma_0^T \bm X) + \varepsilon, \quad \psi_0(\bm x)= x^3, \quad \a_{01} =\a_{02} =\a_{03} = 1/\sqrt{3}, \quad X_1,X_2, X_3 \stackrel{i.i.d}{\sim} N[0,1], \quad  \varepsilon \sim N(0,1),
\end{align}
where $\varepsilon$ is independent of the covariate vector $\bm X = (X_1,X_2, X_3)^T$.

If we would not make the effort to estimate the link function, but just assume that this function is linear, we get, for a sample of size $n=200$, Figure \ref{fig:linear_fit_comparison}(a), where the points represent the values $(\hat\bma_n^T\bmX_i,Y_i)$. Although it might seem somewhat unlikely from Figure \ref{fig:linear_fit_comparison}, this linear regression method produces in fact a $\sqrt{n}$ convergent estimate of $\bma_0$ in the present model.

On the other hand, if we estimate the function $\psi_0$ by the profile Least Squares Estimator (LSE), shown in Figure \ref{fig:linear_fit_comparison}(b), where one uses for fixed $\bma$ the least squares estimate under the restriction that the estimate is nondecreasing in the ordered variables $\bma^T\bm X_i$, and minimize over all such $\bma$, one would hope that the estimate of $\bma_0$ is better because here the link function is estimated rather well. However, as mentioned in Section \ref{sec:intro}, its limiting distribution is unknown, and it is even not known whether it attains the $\sqrt{n}$ rate. This was the motivation to introduce other estimates that are still built on the same pointwise least squares estimate $\hat\psi_{n,\bma}$ for fixed $\bma$, but for which we can prove $\sqrt{n}$ convergence and asymptotic normality.

The crucial model function $u\mapsto\psi_{\bma}(u)=\E\{\bma_0^T\bmX|\bma^T\bmX=u\}$ is in this model given by:
\begin{align*}
\psi_{\bma}(u)=\E\{\bma_0^T\bmX|\bma^T\bmX=u\}
=\frac{\left(\sum_i\a_i\right)u\left\{6\left(\sum_{i}\a_i^2\right)\left(\sum_i\a_i^2-\sum_{i<j}\a_i\a_j\right)+\left(\sum_i\a_i\right)^2u^2\right\}}{3\sqrt{3}\left(\a_1^2+\a_2^2+\a_3^2\right)^3}\,,
\end{align*}
which is clearly differentiable w.r.t.\ $\bma$ and $u$.

\begin{figure}[!h]
	\centering
	\begin{subfigure}{0.3\textwidth}
\includegraphics[width=\textwidth]{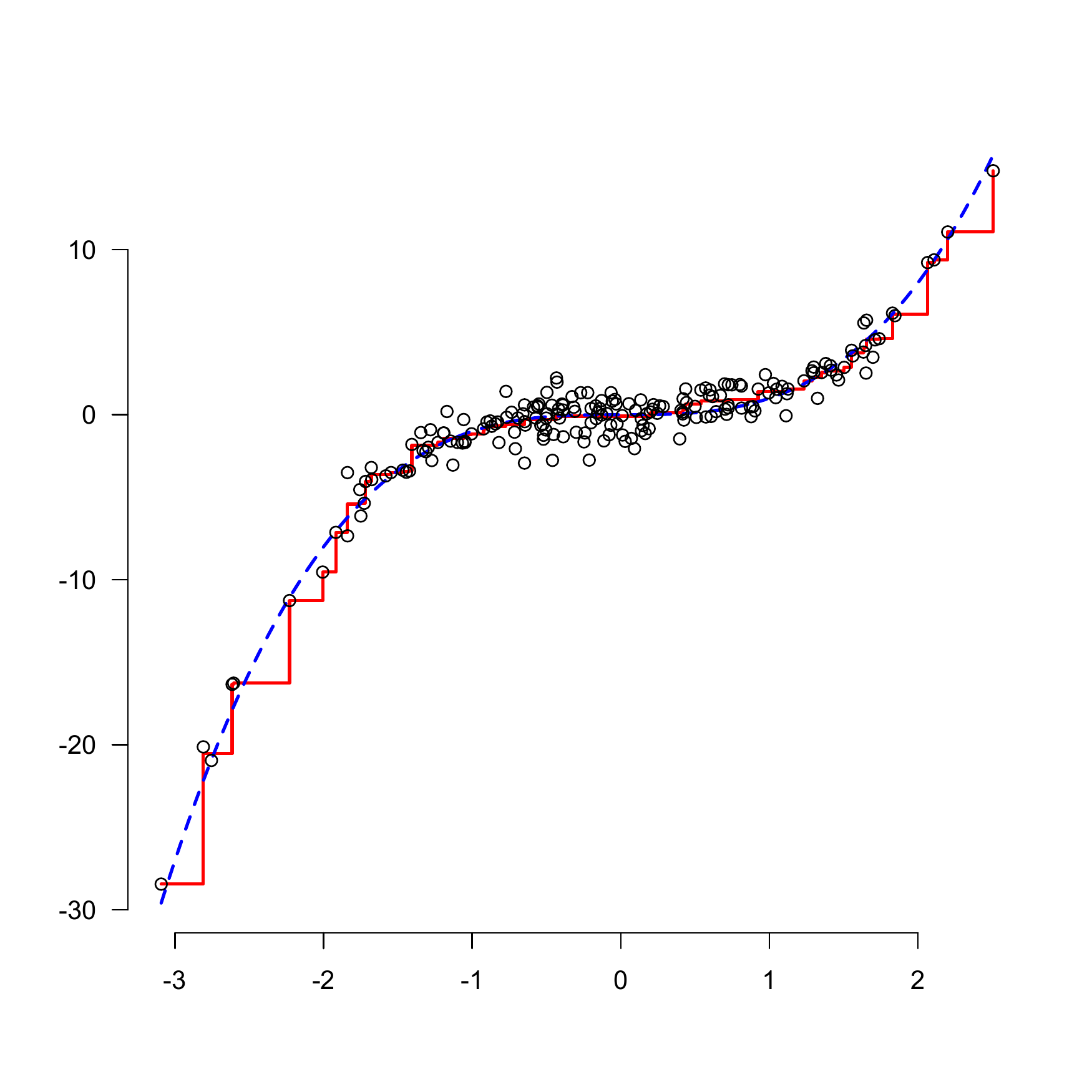}
		\caption{simple score estimate (SSE)}
	\end{subfigure}
	\begin{subfigure}{0.3\textwidth}
	\includegraphics[width=\textwidth]{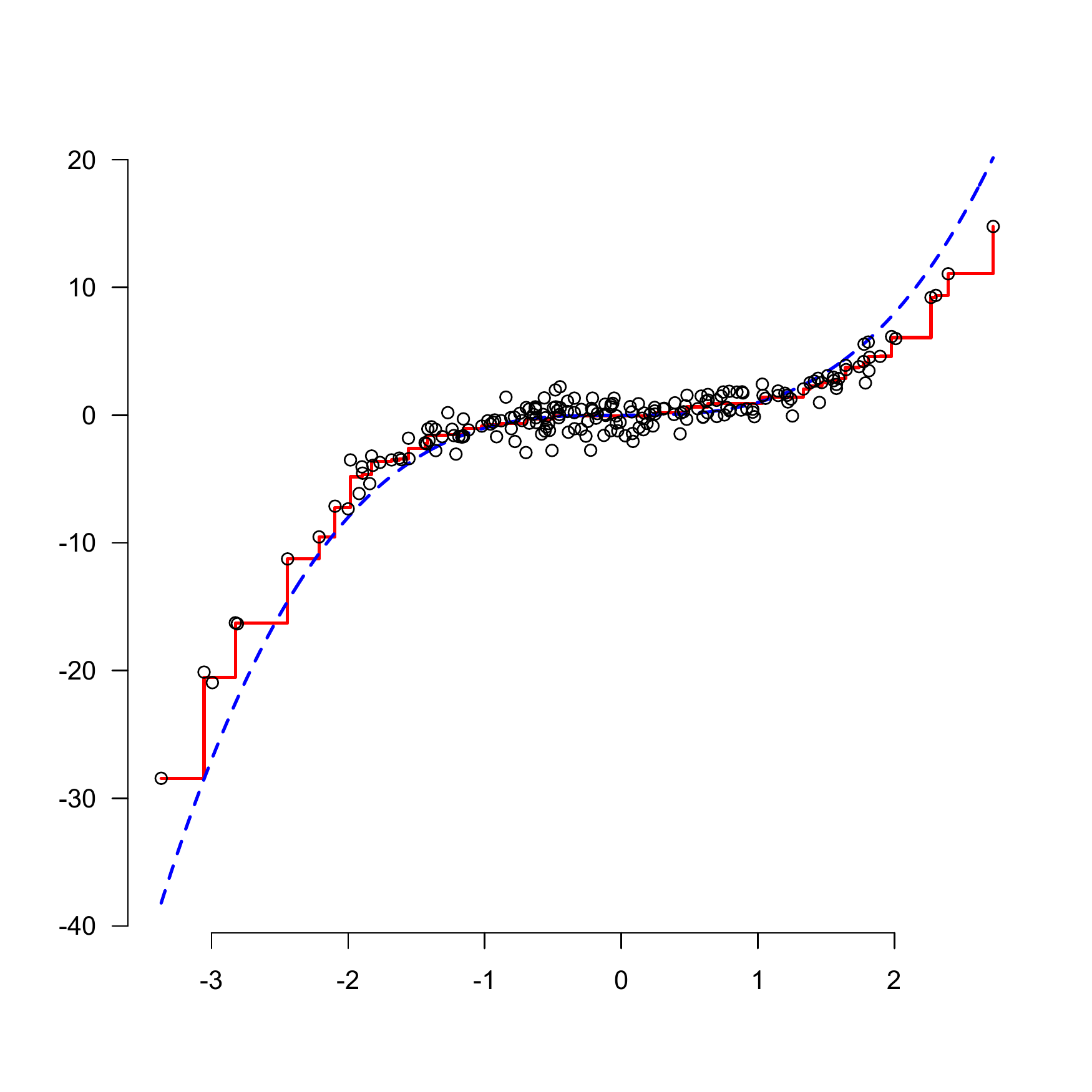}
	\caption{efficient score estimate (ESE)}
\end{subfigure}
	\begin{subfigure}{0.3\textwidth}
		\includegraphics[width=\textwidth]{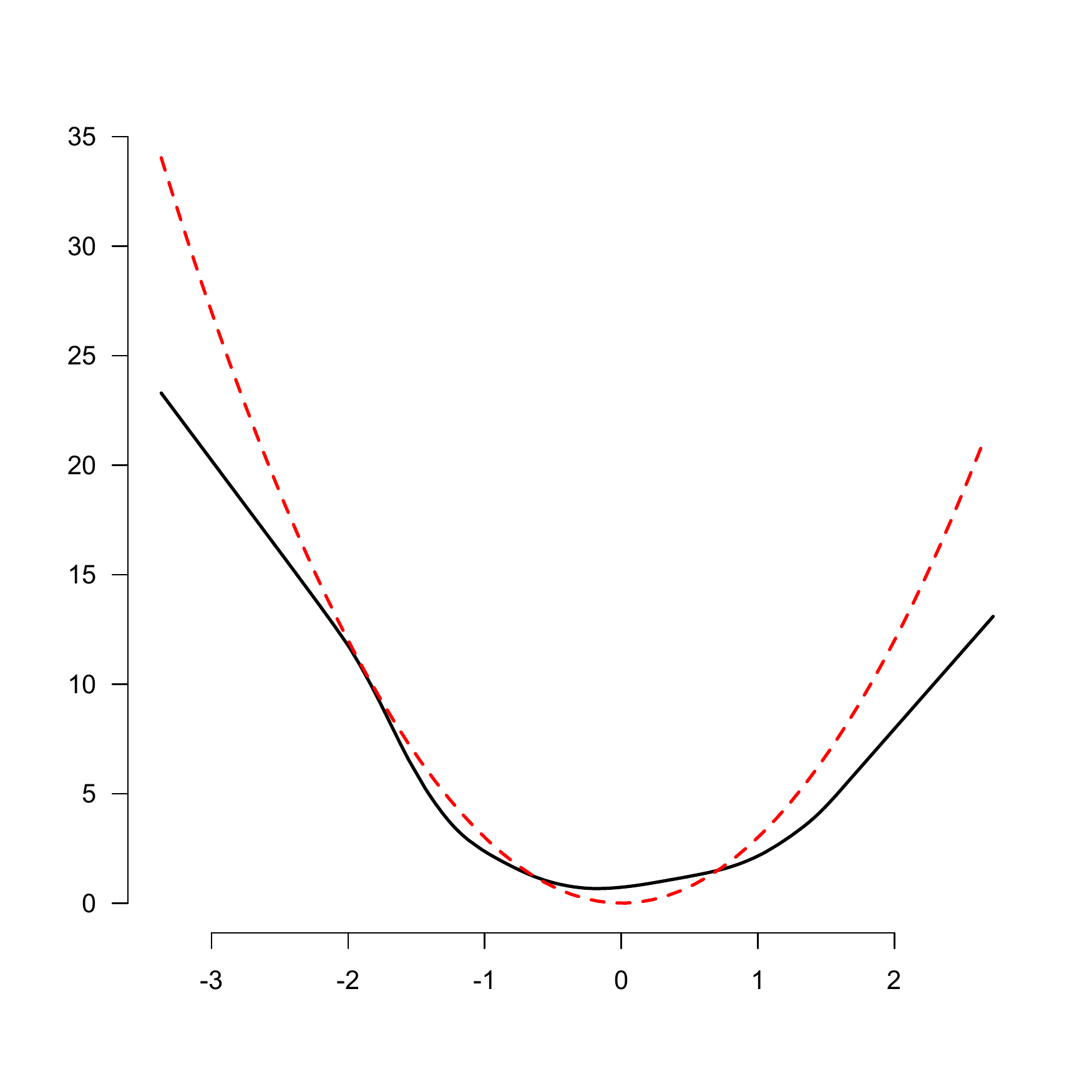}
		\caption{derivative for ESE (\ref{estimate_derivative_psi})}
	\end{subfigure}
\caption{Score estimates of $\psi_0$ ((a) and (b)), and in (c) the derivative (\ref{estimate_derivative_psi}), used in (\ref{ESE_eq}) for computing the efficient score estimate (ESE) (solid curves) for the same sample as in Figure \ref{fig:linear_fit_comparison}. The curves of $\psi_0$ and $\psi_0'$ are dashed.}
\label{derivative_pictures}
\end{figure}

\begin{figure}[!ht]
	\centering
	\begin{subfigure}{0.45\linewidth}
		\includegraphics[width=0.95\textwidth]{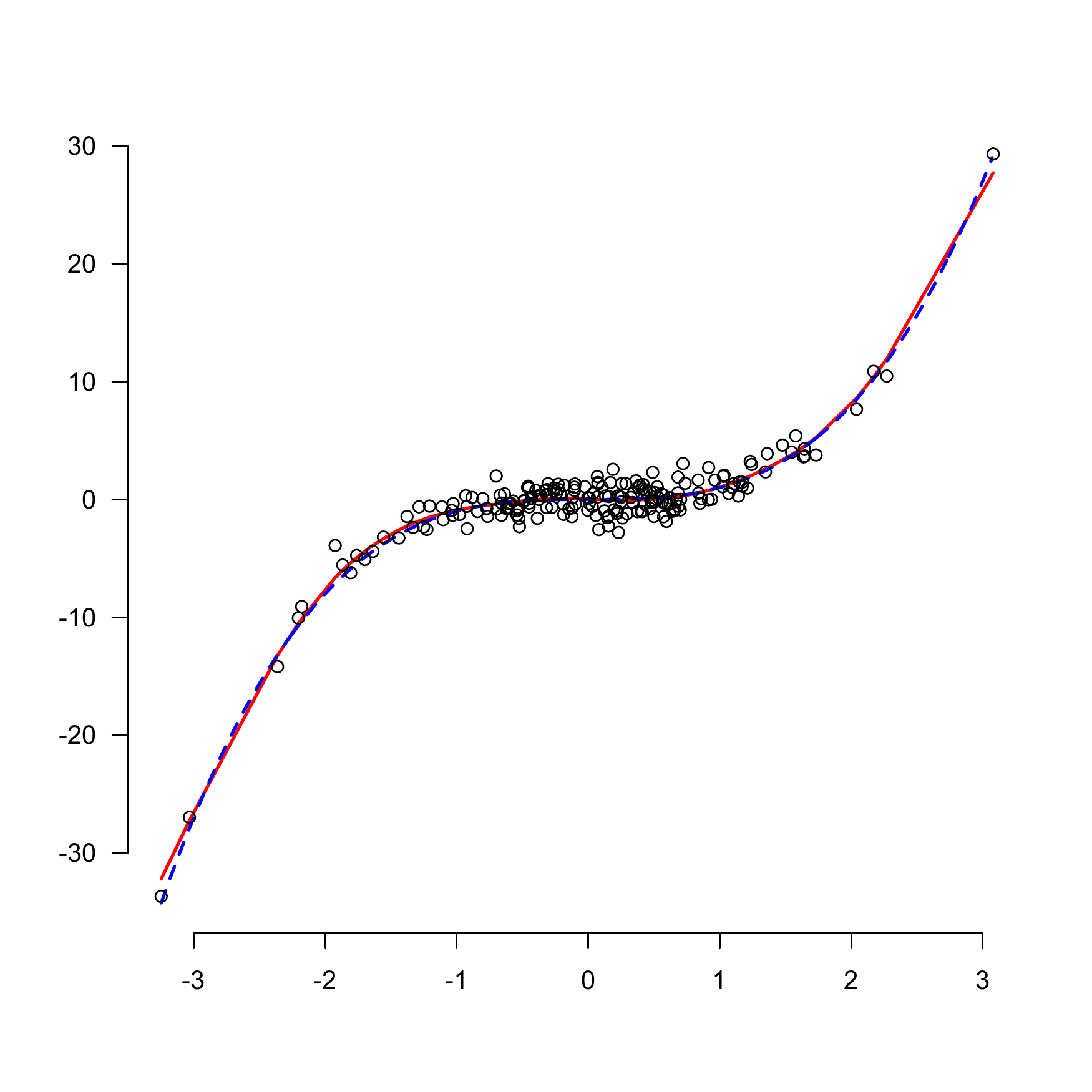}
		\caption{spline estimate PLSE}
	\end{subfigure}
	\begin{subfigure}{0.45\linewidth}
	\includegraphics[width=0.95\textwidth]{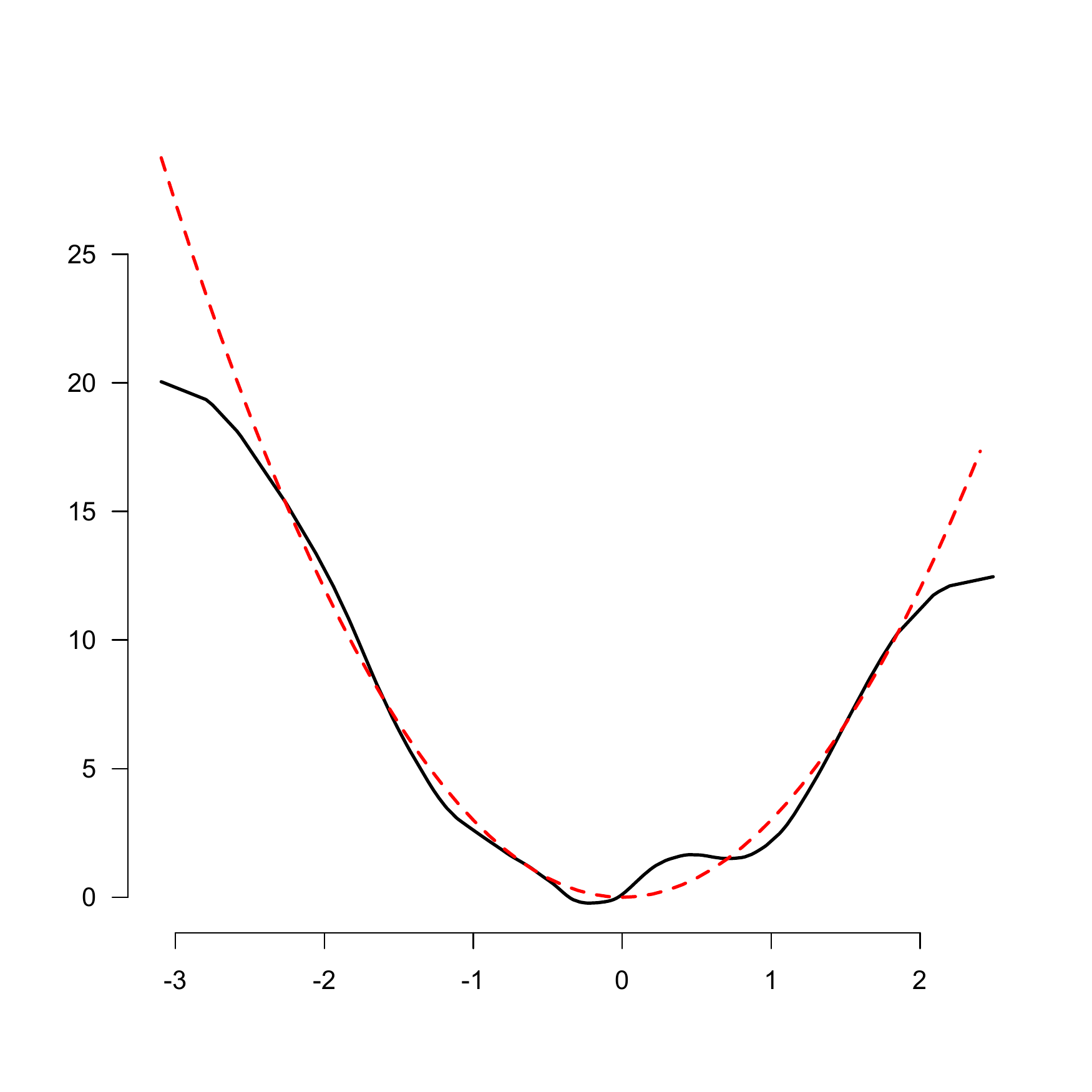}
	\caption{derivative of PLSE}
	\end{subfigure}
	\caption{Cubic spline estimate (a) and its derivative (b) for the same sample as in Figure \ref{derivative_pictures}. Both functions are used in the estimate of $\bma_0$ in an equation of type (\ref{ESE_eq}). The dashed curves are the functions $\psi_0$ and $\psi_0'$, in (a) and (b), respectively.}
	\label{fig:cubic_spline}
\end{figure}

In Section \ref{sec:intro} we defined the Simple Score Estimator (SSE) and the Efficient Score Estimator (ESE), for which pictures of the corresponding functions $\hat\psi_{n,\hat\bma}$ are shown in Figure \ref{derivative_pictures}. For the ESE also the estimate of the derivative, defined by (\ref{estimate_derivative_psi}), is shown in Figure \ref{derivative_pictures} (c). A picture of the Penalized Least Squares Estimate (PLSE) and its derivative, as defined in Section \ref{sec:intro}, is given in Figure \ref{fig:cubic_spline}, where a penalty parameter $\mu=0.1$ is used.

\begin{figure}[!h]
\centering
\includegraphics[width=0.6\textwidth]{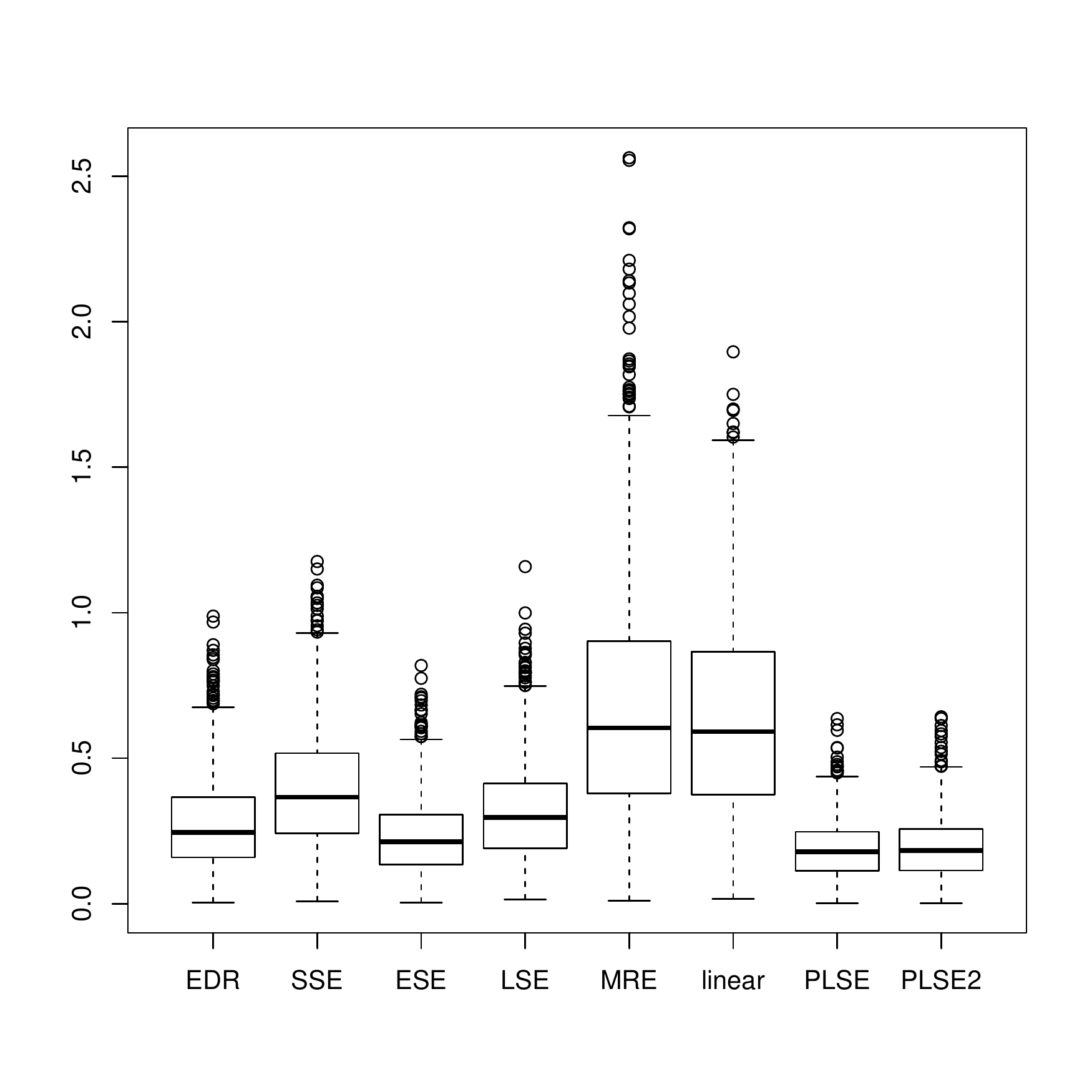}
\caption{Boxplots of $\sqrt{n/d}\,\|\hat\bma_n-\bma_0\|_2$ for $n=200$ and $1000$ replications for the Effective Dimension Reduction estimate (EDR), the Simple Score Estimate (SSE), the Efficient Score Estimate (ESE), the profile Least Squares Estimate (LSE), Han's Maximum Rank Correlation Estimate (MRE), the linear regression estimate and the Penalized Least Squares Estimates PLSE and PLSE2. PLSE is the L-estimate, using (\ref{spline_eq}), whereas PLSE2 is the Penalized Least Squares Estimate, computed by the {\tt R} package {\tt simest}, 
The algorithms for SSE, ESE,  MRE and PLSE and PLSE2 were started at the value of the LSE.  Both the PLSE and the PLSE2 use the penalty parameter $\mu=0.1$.}
\label{boxplot1}
\end{figure}

We compare the behavior of these estimates in the boxplot  Figure \ref{boxplot1} with the behavior of three other estimates, the so-called Effective Dimension Reduction (EDR) method, the linear estimate and Han's maximum correlation estimate. The {\tt R} script for this comparison can  be found on \cite{github:18}.

The simple score estimator (SSE) was computed by minimizing the norm of the left-hand side of (\ref{SSE_eq}),  the efficient score estimator (ESE) by minimizing the norm of the left-hand side of (\ref{ESE_eq}), and the penalized least squares estimator (PLSE) by minimizing the norm of the left-hand side of (\ref{spline_eq}). For the ESE the derivative is estimated by (\ref{estimate_derivative_psi}), for the penalized least squares estimator (PLSE) by the derivative of the cubic spline, which can be computed explicitly, see \cite{green_silverman:94}, Section 2.5.1. For the PLSE we took the penalty parameter $\mu=0.1$. In all cases we get an approximate solution of either equation (\ref{SSE_eq}) (for the SSE) or equation (\ref{ESE_eq}) (ESE) or equation (\ref{spline_eq}) (for the PLSE). The LSE, SSE and ESE estimate $\hat\psi_{n,\bma}$ is, for fixed $\bma$, the isotonic least squares estimate, as defined in Section 3 of \cite{balabdaoui2018}). 

After some experimentation for the best behavior, we used the Hooke-Jeeves method for the PLSE and the Nelder-Mead algorithm for minimizing over $\bma$ in all other cases. The implementation of each method can again be found on \cite{github:18} in the form of an {\tt R} script and {\tt C++} code.

The LSE is computed in the following way. We generate a number of standard normal $d$-dimensional vectors and normalize these to have norm equal to 1. These are taken as initial values of $\bma$ in the search for the minimizing $\hat\bma$. In the example we generated $20$ of such normal vectors for each sample. These were taken as starting values for the search of the minimum of the criterion
\begin{align}
\label{LS_criterion}
n^{-1}\sum_{i=1}^n\left\{Y_i-\hat\psi_{n,\bma}(\bma^T\bmX_i)\right\}^2,
\end{align}
over $\bma$. We do not try to use the constraint $\|\bma\|_2=1$, but just let the Nelder-Mead algorithm look for an $\bma$ minimizing (\ref{LS_criterion}), where we use the following property of the least squares estimate $\hat\psi_{n,\bma}$ for fixed $\bma$:
\begin{align}
\label{rescaling}
\hat\psi_{n,\bma}(\bma^T\bmx)=\hat\psi_{n,c\bma}(c\bma^T\bmx),
\end{align}
for each $c>0$ (only the ordering of the variables $\bma^T\bmX_i$ counts, which does not change if we multiply $\bma$ with $c>0$). So the criterion (\ref{LS_criterion}) also does not change by the scale change (\ref{rescaling}). We normalize the $\hat\alpha$ found for each starting value to have norm $1$. 

Note that the SSE and the ESE use the same estimators as the LSE; they both minimize the criterion (\ref{LS_criterion}) for fixed $\bma$, just as the LSE is doing, yielding the same $\hat\psi_{n,\bma}$. Only the subsequent search for the estimate of $\bma_0$ is done differently: the LSE tries to minimize again, whereas the SSE and ESE try to solve an equation in $\bma$.

The LSE estimate of $\bma_0$ was used as starting value for the search procedures for the SSE, ESE, PLSE and Han's MRE (see \cite{han:87}). The EDR and the linear method do not need a starting value from outside. This is clear for the linear method. For the EDR the starting value is generated within the algorithm by a simple (not $\sqrt{n}$ convergent) average derivative (w.r.t $\bmX$) estimator, which is iteratively improved upon in the algorithm. 

The results of this procedure are shown in Figure \ref{boxplot1}. The EDR was computed using the {\tt R} package {\tt edr}, the PLSE by our own {\tt R} script on \cite{github:18}, and the PLSE2 is computed using the {\tt R} package {\tt simest}, where we also used the LSE as starting value.
It is seen that Han's Maximum Rank Correlation method and the linear regression method have in this case a behavior which is clearly inferior to the other methods. If we add dependence to the distribution of the covariates the linear estimate becomes even worse and the other estimates show roughly the same configuration (not shown here). The profile least squares estimate LSE has a remarkably good behavior, although for this estimator not even the $\sqrt{n}$ convergence has been proved, as in fact has been proved for Han's MRE and the linear regression estimate. The linear regression method attains the $\sqrt{n}$ convergence because the covariates have an elliptically symmetric distribution in the present model, see Section \ref{sec:elliptically_symmetric}.

\section{The linear estimator and elliptically symmetric distributions}
\label{sec:elliptically_symmetric}
It is proved in \cite{Balabdaoui_Durot_Jankowski:2016} that in the case that the distribution of $\bmX$ is elliptically symmetric, ordinary least squares estimates of $\bma_0$ are $\sqrt{n}$ convergent, under some (moment) regularity condition on $\bmX$ and $Y$. Normal distributions are a particular case of elliptically symmetric distributions. So, no matter how far off the linear fit in Figure \ref{fig:linear_fit_comparison} (a) seems, the resulting estimate of $\bma_0$ is $\sqrt{n}$ convergent. We briefly explain this fact, which seems surprising at first sight.

In computing the linear estimate, we first do not attempt to solve for a vector $\bma$ with norm $1$. So we do not use the Lagrange approach, which we used for the SSE, ESE and PLSE, and simply solve the equation
\begin{align}
\label{sample_ellipticity_equation}
\bm S_n\bma=n^{-1}\sum_{i=1}^n\left(\bmX_i-\bar\bmX_n\right)Y_i
\end{align}
in $\bma$, where
\begin{align*}
\bm S_n=n^{-1}\sum_{i=1}^n\left(\bmX_i-\bar{\bmX}_n\right)\left(\bmX_i-\bar{\bmX}_n\right)^T,\qquad \bar{\bmX}_n=n^{-1}\sum_{i=1}^n\bmX_i,
\end{align*}
So the solution in fact minimizes
\begin{align*}
n^{-1}\sum_{i=1}^n\left\{Y_i-\bma^T(\bmX_i-\bar\bmX_n)\right\}^2
\end{align*}
instead of
\begin{align*}
\sum_{i=1}^n\left\{Y_i-\psi_0(\bma^T\bmX_i)\right\}^2.
\end{align*}
Next we claim that the solution $\hat\bma_n$ of the linear problem has the property that $\hat\bma_n/\|\hat\bma_n\|$ is a $\sqrt{n}$ convergent estimate of $\bma_0$.

Let us first investigate why $\hat\bma_n/\|\hat\bma_n\|$ would be consistent. The vector $\hat\bma_n$ converges in probability to a  vector $\tilde\bma$, satisfying
\begin{align}
\label{ellipticity_equation}
\bm\Sigma\tilde\bma=\text{cov}(\bmX,Y),
\end{align}
where $\bm\Sigma$ is the covariance matrix of $\bmX$.
Furthermore,
\begin{align*}
\text{cov}(\bmX,Y)&=\E\{\bmX-\E\bmX)Y=\E\{\bmX-\E\bmX)\psi_0(\bma_0^T\bmX)\\
&=\E\left\{\psi_0(\bma_0^T\bmX)\E\{\bmX-\E\bmX|\bma_0^T\bmX\}\right\}.
\end{align*}
Using the elliptic symmetry, we get:
\begin{align*}
\E\{(\bmX-\E\bmX)|\bma_0^T\bmX\}=\bma_0^T(\bmX-\E\bmX)\,\bm\b,
\end{align*}
for a vector $\bm\b$ such that
\begin{align*}
\E\left\{(\bmX-\E\bmX)\bma_0^T\bmX\right\}=\text{var}(\bma_0^T\bmX)\bm\b=(\bma_0^T\bm\Sigma\bma_0)\bm\b,
\end{align*}
see \cite{Balabdaoui_Durot_Jankowski:2016}. Hence:
\begin{align*}
\bm\b=(\bma_0^T\bm\Sigma\bma_0)^{-1}\bm\Sigma\bma_0,
\end{align*}
and we obtain:
\begin{align*}
\text{cov}(\bmX,Y)=(\bma_0^T\bm\Sigma\bma_0)^{-1}\E\left\{\psi_0(\bma_0^T\bmX)\bma_0^T(\bmX-\E\bmX)\right\}\bm\Sigma\,\bma_0.
\end{align*}

So equation (\ref{ellipticity_equation}) turns into
\begin{align}
\label{ellipticity_equation}
\bm\Sigma\tilde\bma=c\,\bm\Sigma\,\bma_0,
\end{align}
where
\begin{align}
\label{covariance_constant}
c=(\bma_0^T\bm\Sigma\bma_0)^{-1}\text{cov}\left(\psi_0(\bma_0^T\bmX),\bma_0^T\bmX\right),
\end{align}
showing that $\bm\Sigma\tilde\bma$ is a multiple of $\bm\Sigma\bma_0$. Assuming $\bm\Sigma$ to be non-singular, we therefore get that $\tilde\bma$ is a multiple of $\bma_0$. This means that $\hat\bma_n/\|\hat\bma_n\|$ converge in probability to $\bma_0$, since we may assume in the present case that $\text{cov}\left(\psi_0(\bma_0^T\bmX),\bma_0^T\bmX\right)>0$, using the fact that $\psi_0$ is a nondecreasing function.

To prove the $\sqrt{n}$ convergence and asymptotic normality, we first derive the asymptotic distribution of $\tilde\bma_n$, solving  (\ref{sample_ellipticity_equation}). We have:
\begin{align*}
\tilde\bma_n=\bm S_n^{-1}\frac1n\sum_{i=1}^n\left(\bmX_i-\bar\bmX_n\right)Y_i.
\end{align*}
A simple application of the central limit theorem, together with Slutsky's theorem, give that
\begin{align*}
\sqrt{n}\left\{\tilde\bma_n-c\,\bma_0\right\}\stackrel{d}\longrightarrow N(\bm0,\bm A),
\end{align*}
where $c$ is defined by (\ref{covariance_constant}) and $\bm A$ is defined by
\begin{align}
\label{def_Gamma}
\bm A=\bm\Sigma^{-1}\bm\Gamma\bm\Sigma^{-1},\qquad \bm\Gamma=\E\left(Y^2\bmX\bmX^T\right)-(\E Y\bmX)(\E Y\bmX)^T,
\end{align}
where we follow the notation in \cite{Balabdaoui_Durot_Jankowski:2016}. Another application of Slutsky's theorem now yields that
\begin{align*}
\sqrt{n}\left\{\tilde\bma_n/\|\tilde\bma_n\|-\bma_0\right\}\longrightarrow N(\bm0,\bm V),
\end{align*}
where
\begin{align*}
\bm V=c^{-2}\left\{\bm I-\bma_0\bma_0^T\right\}\bm\Sigma^{-1}\bm\Gamma\bm\Sigma^{-1}\left\{\bm I-\bma_0\bma_0^T\right\}.
\end{align*}
Details on the Jacobian for the last application of Slutsky's theorem are given in  \cite{Balabdaoui_Durot_Jankowski:2016}.

The theorem in  \cite{Balabdaoui_Durot_Jankowski:2016} runs as follows.

\begin{theorem}
\label{Th:H-LFLSE}
Let $(\bmX_1,Y_1),\dots,(\bm X_n,Y_n)$ be an i.i.d.\ sample from $(\bmX,Y)$ such that $E(Y|\bmX)=\psi_0(\bma_0\bmX)$ almost surely, where $\psi_0$ is non-decreasing and $\bma_0^T\bma_0=1$. Suppose that $\bm X$ has an elliptically symmetric distribution with finite mean $\bm\mu\in\R^d$ and a positive definite covariance matrix $\bm\Sigma$. Assume, moreover, that $\E\|Y\bmX\|<\infty$ and that there exists a nonempty interval $[a,b]$ on which $\psi_0$ is strictly increasing. Then, as $n\to\infty$, the estimator $\tilde{\bma}_n=\hat{\bm\alpha}_n/\|\hat{\bm\alpha}_n\|$, where $\hat\bma_n$ is defined by
\begin{align}
\label{LS_BDJ}
\hat{\bm\alpha}_n=\text{\rm argmin}_{\bm\a\in\R^d}\sum_{i=1}^n\left\{Y_i-\bm\a^T(\bm X_i-\bar{\bm X}_n)\right\}^2,
\end{align}
converges in probability to $\bma_0$. If, moreover, $\E Y^2\|\bmX\|^2<\infty$, $\sqrt{n}\{\tilde\bma_n-\bma_0\}$ converges in distribution to a normal distribution with mean $\bm 0$ and covariance matrix
\begin{align*}
\frac1{c^2}\left(\bm I-\bm\a_0\bm\a_0^T\right)\bm\Sigma^{-1}\bm \Gamma\bm\Sigma^{-1}\left(\bm I-\bm\a_0\bm\a_0^T\right),\qquad
c=\text{\rm Cov}\left( \psi_0(\bma_0^T\bm X), \bma_0^T\bmX  \right)/\bma_0^T\bm \Sigma\bma_0,
\end{align*}
where $\bm \Gamma$ is defined by (\ref{def_Gamma}).
\end{theorem}

\section{The (profile) LSE}
\label{section:LSE}
The consistency of the profile LSE can be derived in the following way.  The function $\bmx\mapsto\hat\psi_{n,\hat\bma_n}(\hat\bma_n^T\bmx)$ minimizes the expression:
\begin{align*}
&n^{-1}\sum_{i=1}^n\left\{\psi_{n,\bma}(\bma^TX_i)-Y_i\right\}^2-n^{-1}\sum_{i=1}^n\left\{\psi_0(\bma_0^TX_i)-Y_i\right\}^2\\
&=n^{-1}\sum_{i=1}^n\left\{\psi_{n,\bma}(\bma^TX_i)^2-\psi_0(\bma_0^TX_i)^2\right\}-2n^{-1}\sum_{i=1}^n\left\{\psi_{n,\bma}(\bma^TX_i)-\psi_0(\bma_0^TX_i)\right\}Y_i\\
&=n^{-1}\sum_{i=1}^n\left\{\psi_{n,\bma}(\bma^TX_i)-\psi_0(\bma_0^TX_i)\right\}^2
-2n^{-1}\sum_{i=1}^n\left\{\psi_{n,\bma}(\bma^TX_i)-\psi_0(\bma_0^TX_i)\right\}\left\{Y_i-\psi_0(\bma_0^T\bmX_i\right\},
\end{align*}
under the restriction $\|\bma\|_2=1$, where we subtract in the first line a term that does not influence the minimization.
Since  the function $\bmx\mapsto\hat\psi_{n,\hat\bma_n}(\hat\bma_n^T\bmx)$ minimizes this expression, it gives in particular a smaller value of the expression than the function $\bmx\mapsto\psi_0(\bma_0^T\bmx)$. This implies:
\begin{align*}
n^{-1}\sum_{i=1}^n\left\{\hat\psi_{n,\hat\bma_n}(\hat\bma_n^T\bmX_i)-\psi_0(\bma_0^TX_i)\right\}^2
- 2n^{-1}\sum_{i=1}^n\left\{\hat\psi_{n,\hat\bma_n}(\hat\bma_n^T\bmX_i)-\psi_0(\bma_0^TX_i)\right\}\left\{Y_i-\psi_0(\bma_0^T\bmX_i\right\}\le0.
\end{align*}
and hence
\begin{align*}
n^{-1}\sum_{i=1}^n\left\{\hat\psi_{n,\hat\bma_n}(\hat\bma_n^T\bmX_i)-\psi_0(\bma_0^TX_i)\right\}^2
\le 2n^{-1}\sum_{i=1}^n\left\{\hat\psi_{n,\hat\bma_n}(\hat\bma_n^T\bmX_i)-\psi_0(\bma_0^TX_i)\right\}\left\{Y_i-\psi_0(\bma_0^T\bmX_i\right\},
\end{align*}
So we find:
\begin{align*}
\limsup_{k\to\infty}{n}^{-1}\sum_{i=1}^{n}\left\{\hat\psi_{n,\hat\bma_{n}}(\hat\bma_{n}^T\bmX_i)-\psi_0(\bma_0^T\bmX_i)\right\}^2\le0,
\end{align*}
almost surely, which implies
\begin{align*}
\lim_{n\to\infty}n^{-1}\sum_{i=1}^n\left\{\hat\psi_{n,\hat\bma_n}(\hat\bma_n^T\bmX_i)-\psi_0(\bma_0^T\bmX_i)\right\}^2=0.
\end{align*}
almost surely. From this and the fact that $\bma_n$ has norm 1, the consistency of $\hat\bma_n$ for $\bma_0$ can easily be deduced.

The  convergence rate and asymptotic distribution of the LSE are unknown. In \cite{Balabdaoui_Durot_Jankowski:2016} a cube root convergence rate for both $\hat\psi_{n,\hat\bma_n}$ and $\hat\bma_n$ is derived.

\section{Consistency and asymptotic normality of the SSE}
\label{sec:consistencySSE}
We prove consistency under the conditions of \cite{balabdaoui2018}.
Let $(\hat\bma_{n_k})$ be a subsequence converging to a vector $\tilde\bma$, satisfying $\|\tilde\bma\|_2=1$. The vector $\tilde\bma$ has to satisfy:
\begin{align*}
\left(\bmI-\tilde\bma\tilde\bma^T\right)\E\left[\bmX\left\{Y-\psi_{\tilde\bma}(\tilde\bma^T\bmX)\right\}\right]=\bm0,
\end{align*}
where
\begin{align*}
\psi_{\tilde\bma}(u)=\E\left\{\psi_0(\bma_0^T\bmX)|\tilde\bma^T\bmX=u\right\}.
\end{align*}
Since $\E\left\{Y|\bmX\right\}=\psi_0(\bma_0^T\bmX)$, we can write this in the form
\begin{align}
\label{asymptotic_score_eq}
&\left(\bmI-\tilde\bma\tilde\bma^T\right)\E\left[\bmX\left\{\psi_0(\bma_0^T\bmX)-\E\left\{\psi_0(\bma_0^T\bmX)|\tilde\bma^T\bmX\right\}\right\}\right]\nonumber\\
&=\left(\bmI-\tilde\bma\tilde\bma^T\right)\E\left[\text{cov}\left(\bmX,\psi_0(\bma_0^T\bmX)\right\}\tilde\bma^T\bmX\right]]\nonumber\\
&=\bm0.
\end{align}
Premultiplying this by $(\bma_0-\tilde\bma)^T$ we get:
\begin{align}
\label{asymptotic_score_eq2}
&(\bma_0-\tilde\bma)^T\left(\bmI-\tilde\bma\tilde\bma^T\right)\E\left[\text{cov}\left(\bmX,\psi_0(\bma_0^T\bmX)\right\}\tilde\bma^T\bmX\right]]\nonumber\\
&=\E\left[\text{cov}\left((\bma_0-\tilde\bma)^T\bmX-(\bma_0-\tilde\bma)^T\tilde\bma\left(\tilde\bma^T\bmX\right),
\psi_0(\tilde\bma^T\bmX+(\bma_0-\tilde\bma)^T\bmX)\right)\bigm|\tilde\bma^T\bmX\right]\nonumber\\
&=\E\left[\text{cov}\left((\bma_0-\tilde\bma)^T\bmX,
\psi_0(\tilde\bma^T\bmX+(\bma_0-\tilde\bma)^T\bmX)\right)\bigm|\tilde\bma^T\bmX\right]=0.
\end{align}
So either $\tilde\bma=\pm\bma_0$ (using $\|\tilde\a\|_2=1$), or $\tilde\bma_0$ has a component in the space perpendicular to  $\bma$, in which case we get that $(\bma_0-\tilde\bma)^T\bmX$ is, conditionally on $\tilde\bma^T\bmX$, nondegenerate and that
\begin{align*}
\E\left[\text{cov}\left((\bma_0-\tilde\bma)^T\bmX,
\psi_0(\tilde\bma^T\bmX+(\bma_0-\tilde\bma)^T\bmX)\right)\bigm|\tilde\bma^T\bmX\right]>0.
\end{align*}
contradicting (\ref{asymptotic_score_eq2}). So we obtain
$$
\tilde\bma=\pm\bma_0.
$$
Since a sufficiently small neighborhood around $\bma_0$ excludes $-\bma_0$, the consistency follows if we only look for a solution in this neighborhood.

We now give the argument for the following theorem on the asymptotic normality of $\sqrt{n}\{\hat\bma_n-\bma_0\}$.

\begin{theorem}
	\label{theorem:asymptotics}
	Let Assumptions A1-A6 of \cite{balabdaoui2018}) be satisfied and assume that $\|\bma_0\|_2=1$ and that
	\begin{description}
		\item A7. There exists a $\d_0>0$ such that for all $\bma \ne \bma_0$, satisfying $\|\bma-\bma_0\|<\d_0$ and $\|\bma\|_2=1$,
		\begin{align*}
\E\left[\text{\rm cov}\left((\bma_0-\tilde\bma)^T\bmX,
\psi_0(\bma_0^T\bmX)\right)\bigm|\bma^T\bmX\right]>0.
\end{align*}
\item A9. $ \E\Bigl[\psi_0'(\bma_0^T\bm X)\,\text{\rm  var}(\bm X|\bma_0^T\bm X)\Bigr]$  has rank $d-1$.
	\end{description}
	 
	 Let  $\hat \bma_n$ be defined as a crossing of zero of the left-hand side of (\ref{SSE_eq}). 	Let the matrices $\bm A$ and $\bm\Sigma$ be defined by:
		\begin{align}
		\label{def_A}
		\bm A:=\E\Bigl[\psi_0'(\bma_0^T\bm X)\,\text{\rm Cov}(\bm X|\bm\a_0^T\bm X)\Bigr],
		\end{align}
		and
		\begin{align}
		\label{def_Sigma}
		\bm \Sigma:=\E\left[\left\{Y -\psi_0(\bm\a_0^T\bm X)\right\}^2\,\left\{\bm X -\E(\bm X|\bm\a_0^T\bm X) \right\}\left\{\bm X -\E(\bm X|\bm\a_0^T\bm X) \right\}^T\right].
		\end{align}
		Then 
		\begin{align*}
		\sqrt n (\hat \bma_n - \bma_0) \to_d N_{d}\left(\bm 0,  \bm A^- \bm \Sigma \bm A^-\right),
		\end{align*}
		where $\bm A^{-}$ is the Moore-Penrose inverse of $\bm A$ and $N_d$ denotes the degenerate $d$-dimensional normal distribution with mean zero and covariance matrix $\bm A^- \bm \Sigma \bm A^-$.
\end{theorem}

\begin{remark}
{\rm This is a modification of the asymptotic normality part of Theorem 5 in \cite{balabdaoui2018}, avoiding the manipulations with the Jacobians, which were needed there because of the lower dimensional parametrization.
}
\end{remark}

Before going into the proof, we highlight the essential elements. We first note that we may assume, as in \cite{balabdaoui2018}, that the estimator $\hat\bma_n$ satisfies the equation:
\begin{align}
\label{simple_score_eq}
\left(\bmI-\hat\bma_n\hat\bma_n^T\right)\int\bmx\left\{y-\hat\psi_{n,\hat\bma_n}(\hat\bma_n^T\bmx)\right\}\,d\P_n(\bmx,y)=\bm0.
\end{align}
Using the characterization of the least squares estimator $\hat\psi_{n,\bma}$ for fixed $\bma$, we show that
\begin{align*}
&\int\bmx\left\{y-\hat\psi_{n,\hat\bma_n}(\hat\bma_n^T\bmx)\right\}\,d\P_n(\bmx,y)\\
&=\int\left\{\bmx-\E(X|\hat\bma_n^T\bmX)\right\}\left\{y-\hat\psi_{n,\hat\bma_n}(\hat\bma_n^T\bmx)\right\}\,d\P_n(\bmx,y)
+o_p(n^{-1/2})+o_p(\hat\bma_n-\bma_0),
\end{align*}
see (\ref{fundamental_relation_SSE}) below. Next we show that this, together with (\ref{simple_score_eq}), leads to the asymptotic equation
\begin{align*}
&\E\left\{\psi_0'(\bma_0^T\bmX)\left\{\bmX-\E\left(\bm X|\bma_0^T\bm x\right)\right\}\left\{\bmX-\E\left(\bm X|\bma_0^T\bm X\right)\right\}^T\right\}\left\{\hat\bma_n-\bma_0\right\}\\
&=\int\left\{\bmx-\E\left(\bm X|\bma_0^T\bm x\right)\right\}\left\{y-\psi_0(\bma_0^T\bmx)\right\}
\,d\bigl(\P_n-P_0\bigr)(\bmx,y)
+o_p\left(\hat\bma_n-\bma_0\right)+o_p\left(n^{-1/2}\right).
\end{align*}
see (\ref{SSE_equation}).
The dominating terms on both sides are perpendicular to $\bma_0$. From this, the consistency and the relation $\|\bma\|_2=\|\bma_0\|_2=1$, the result follows.

\begin{proof}
We define the piecewise constant function $\bar E_{n, \bma}$
\begin{eqnarray}
\label{E_n-def}
	\bar E_{n, \bma}(u)  =  \left \{
	\begin{array}{lll}
		\E\left[\bm X| \bma^T\bm X= \t_{i,\bmb}\right] \ \  \ \ \ \ \ \  \ \text{ if $\psi_{\bma}(u)  > \hat\psi_{n\bma}(\tau_i)$  \ for all $u \in (\tau_i, \tau_{i+1})$}, \\
		\E\left[\bm X| \bma^T\bm X= s\right] \ \ \ \  \ \  \  \ \ \ \  \ \text{ if $\psi_{\bma}(s)  = \hat\psi_{n\bma}(s)$ \ for some $s \in (\tau_i, \tau_{i+1})$}, \\
		\E\left[\bm X| \bma^T\bm X= \t_{i+1,\bmb}\right]\ \ \ \ \ \ \ \text{if $\psi_{\bma}(u) < \hat\psi_{n\bma}(\tau_i)$  \ for all $u \in (\tau_i, \tau_{i+1})$},
	\end{array}
	\right.
\end{eqnarray}
where the $\t_i$ are the ordered points of jump of the least squares estimate $\hat\psi_{n\bma}$, and where $\psi_{\bma}(u)= \E\left\{\psi_0(\bma_0^T\bmx)|\bma^T\bmx=u\right\}$. The origin of this method is the treatment of smooth functionals in the interval censoring model, see, e.g., p.\ 290 of \cite{piet_geurt:14}.

Then, since $\hat\psi_{n,\bma}$ is the least squares estimator for fixed $\bma$ and since the function $\bar E_{n, \bma} $ is constant between jumps:
\begin{align*}
\int\bar E_{n,\hat\bma_n}(\hat\bma_n^T\bmx)\left\{y-\hat\psi_{n\hat\bma_n}(\hat\bma_n^T\bmx)\right\}\,d\P_n(\bmx,y)=\bm0,
\end{align*}
implying that we can write (\ref{simple_score_eq}) in the form:
\begin{align*}
&\left(\bmI-\hat\bma_n\hat\bma_n^T\right)\int\left\{\bmx-\bar E_{n,\hat\bma_n}(\hat\bma_n^T\bmx)\right\}\left\{y-\hat\psi_{n\hat\bma_n}(\hat\bma_n^T\bmx)\right\}\,d\P_n(\bmx,y)
=\bm0.
\end{align*}
Note that $\bar E_{n, \bma}$ cannot computed in practice, since we cannot compute the function $\psi_{\bma}$ using the data only.

We have:
\begin{align*}
&\int\left\{\bmx-\bar E_{n,\hat\bma_n}(\hat\bma_n^T\bmx)\right\}\left\{y-\hat\psi_{n\hat\bma_n}(\hat\bma_n^T\bmx)\right\}\,d\P_n(\bmx,y)\\
&=\int\left\{\bmx-\E\left(\bm X|\hat\bma_n^T\bm x\right)\right\}\left\{y-\hat\psi_{n\hat\bma_n}(\hat\bma_n^T\bmx)\right\}\,d\P_n(\bmx,y)\\
&\qquad+\int\left\{\E\left(\bm X|\hat\bma_n^T\bm x\right)-\bar E_{n,\hat\bma_n}(\hat\bma_n^T\bmx)\right\}\left\{y-\hat\psi_{n\hat\bma_n}(\hat\bma_n^T\bmx)\right\}\,d\P_n(\bmx,y).
\end{align*}
The last term can be written:
\begin{align*}
&\int\left\{\E\left(\bm X|\hat\bma_n^T\bm x\right)-\bar E_{n,\hat\bma_n}(\hat\bma_n^T\bmx)\right\}\left\{y-\hat\psi_{n\hat\bma_n}(\hat\bma_n^T\bmx)\right\}\,d\left(P_n-P_0\right)(\bmx,y)\\
&+\int\left\{\E\left(\bm X|\hat\bma_n^T\bm x\right)-\bar E_{n,\hat\bma_n}(\hat\bma_n^T\bmx)\right\}\left\{\psi_0(\bma_0^T\bmx)-\psi_{\hat\bma_n}(\hat\bma_n^T\bmx)\right\}\,dP_0(\bmx,y)\\
&+\int\left\{\E\left(\bm X|\hat\bma_n^T\bm x\right)-\bar E_{n,\hat\bma_n}(\hat\bma_n^T\bmx)\right\}\left\{\psi_{\hat\bma_n}(\hat\bma_n^T\bmx)-\hat\psi_{n,\hat\bma_n}(\hat\bma_n^T\bmx)\right\}\,dP_0(\bmx,y)
\end{align*}
Since, by the choice of $\bar E_{n, \hat\bma_n}$, 
\begin{align}
\label{bound_on_interval}
\left\|\E\left(\bm X|\hat\bma_n^T\bm x\right)-\bar E_{n,\hat\bma_n}(\hat\bma_n^T\bmx)\right\|\le c|\psi_{n\hat\bma_n}(\hat\bma_n^T\bmx)-\psi_{\hat\bma_n}(\hat\bma_n^T\bmx)|,
\end{align}
(see for this type of argument for example the top of p.\ 308 in \cite{piet_geurt:14}), we find that the first integral is $o_p(n^{-1/2})$.
Using (\ref{bound_on_interval}) again, together with the fact that $|\psi_0(\bma_0^T\bmx)-\psi_{\hat\bma_n}(\hat\bma_n^T\bmx)|$ is bounded by a constant times $\|\hat\bma_n-\bma_0\|$, we find that the second term is of smaller order than  $\|\hat\bma_n-\bma_0\|$.
Finally, again using (\ref{bound_on_interval}) we find from the Cauchy-Schwarz inequality that the third term is bounded above by a constant times
\begin{align*}
\int\left\{\psi_{\hat\bma_n}(\hat\bma_n^T\bmx)-\hat\psi_{n,\hat\bma_n}(\hat\bma_n^T\bmx)\right\}^2\,dP_0(\bmx,y)=O_p\left(n^{-2/3}\right).
\end{align*}

So we get the relation
\begin{align}
\label{fundamental_relation_SSE}
&\int\bmx\left\{y-\hat\psi_{n\hat\bma_n}(\hat\bma_n^T\bmx)\right\}\,d\P_n(\bmx,y)\nonumber\\
&=\int\left\{\bmx-\E\left(\bm X|\hat\bma_n^T\bm x\right)\right\}\left\{y-\hat\psi_{n\hat\bma_n}(\hat\bma_n^T\bmx)\right\}\,d\P_n(\bmx,y)
+o_p(n^{-1/2})+o_p(\hat\bma_n-\bma_0).
\end{align}
Furthermore,
\begin{align*}
&\int\left\{\bmx-\E\left(\bm X|\hat\bma_n^T\bm x\right)\right\}\left\{y-\hat\psi_{n\hat\bma_n}(\hat\bma_n^T\bmx)\right\}\,d\P_n(\bmx,y)\\
&=\int\left\{\bmx-\E\left(\bm X|\hat\bma_n^T\bm x\right)\right\}\left\{y-\hat\psi_{n\hat\bma_n}(\hat\bma_n^T\bmx)\right\}\,dP_0(\bmx,y)\\
&\qquad+\int\left\{\bmx-\E\left(\bm X|\hat\bma_n^T\bm x\right)\right\}\left\{y-\hat\psi_{n\hat\bma_n}(\hat\bma_n^T\bmx)\right\}
\,d\bigl(\P_n-P_0\bigr)(\bmx,y)\\
&\int\left\{\bmx-\E\left(\bm X|\hat\bma_n^T\bm x\right)\right\}\left\{y-\psi_{\hat\bma_n}(\hat\bma_n^T\bmx)\right\}\,dP_0(\bmx,y)\\
&\qquad+\int\left\{\bmx-\E\left(\bm X|\hat\bma_n^T\bm x\right)\right\}\left\{\psi_{\hat\bma_n}(\hat\bma_n^T\bmx)-\hat\psi_{n\hat\bma_n}(\hat\bma_n^T\bmx)\right\}\,dP_0(\bmx,y)\\
&\qquad+\int\left\{\bmx-\E\left(\bm X|\hat\bma_n^T\bm x\right)\right\}\left\{y-\hat\psi_{n\hat\bma_n}(\hat\bma_n^T\bmx)\right\}
\,d\bigl(\P_n-P_0\bigr)(\bmx,y)\\
&=\int\left\{\bmx-\E\left(\bm X|\hat\bma_n^T\bm x\right)\right\}\left\{\psi_0(\bma_0^T\bmx)-\psi_{\hat\bma_n}(\hat\bma_n^T\bmx)\right\}\,dP_0(\bmx,y)\\
&\qquad+\int\left\{\bmx-\E\left(\bm X|\hat\bma_n^T\bm x\right)\right\}\left\{y-\hat\psi_{n\hat\bma_n}(\hat\bma_n^T\bmx)\right\}
\,d\bigl(\P_n-P_0\bigr)(\bmx,y),
\end{align*}
where the last equality follows from:
\begin{align*}
&\int\left\{\bmx-\E\left(\bm X|\hat\bma_n^T\bm x\right)\right\}\left\{\psi_{\hat\bma_n}(\hat\bma_n^T\bmx)-\hat\psi_{n\hat\bma_n}(\hat\bma_n^T\bmx)\right\}\,dP_0(\bmx,y)\\
&=\E\left[\E\left\{\left\{\bmX-\E\left(\bm X|\hat\bma_n^T\bm X\right)\right\}\bigl\{\psi_{\hat\bma_n}(\hat\bma_n^T\bmX)-\hat\psi_{n\hat\bma_n}(\hat\bma_n^T\bmX)\bigr\}\Bigm|\hat\bma_n^T\bmX\right\}\right]\\
&=0.
\end{align*}
So, using
\begin{align*}
\psi_0(\bma_0^T\bmx)-\psi_{\hat\bma_n}(\hat\bma_n^T\bmx)=-\psi_0'(\bma_0^T\bmX)\left\{\bmX-\E\left(\bm X|\bma_0^T\bm X\right)\right\}^T\left\{\hat\bma_n-\bma_0\right\}+o_p\left(\hat\bma_n-\bma_0\right),
\end{align*} 
we end up with the equation
\begin{align}
\label{simple_score_eq2}
&\int\left\{\bmx-\E\left(\bm X|\hat\bma_n^T\bm x\right)\right\}\left\{\psi_0(\bma_0^T\bmx)-\psi_{\hat\bma_n}(\hat\bma_n^T\bmx)\right\}\,dP_0(\bmx,y)\nonumber\\
&=\E\left\{\bmX-\E\left(\bm X|\hat\bma_n^T\bm x\right)\right\}\psi_0'(\bma_0^T\bmX)\left\{\bmX-\E\left(\bm X|\bma_0^T\bm X\right)\right\}^T\left\{\hat\bma_n-\bma_0\right\}\nonumber\\
&=\int\left\{\bmx-\E\left(\bm X|\hat\bma_n^T\bm x\right)\right\}\left\{y-\hat\psi_{n\hat\bma_n}(\hat\bma_n^T\bmx)\right\}
\,d\bigl(\P_n-P_0\bigr)(\bmx,y)+o_p\left(\hat\bma_n-\bma_0\right)+o_p(n^{-1/2}),
\end{align}
where we use that premultiplying with $\hat\bma_n\hat\bma_n^T$ yields zero for the dominating terms.

Using $\E(\bmX|\hat\bma_n^T\bmX)-\E\left(\bm X|\bma_0^T\bmX\right)=O_p(\hat\bma_n-\bma_0)$, it is easily seen that
\begin{align*}
&\E\left\{\bmX-\E\left(\bm X|\hat\bma_n^T\bm x\right)\right\}\psi_0'(\bma_0^T\bmX)\left\{\bmX-\E\left(\bm X|\bma_0^T\bm X\right)\right\}^T\left\{\hat\bma_n-\bma_0\right\}\\
&=\E\left\{\bmX-\E\left(\bm X|\bma_0^T\bm x\right)\right\}\psi_0'(\bma_0^T\bmX)\left\{\bmX-\E\left(\bm X|\bma_0^T\bm X\right)\right\}^T\left\{\hat\bma_n-\bma_0\right\}+o_p\left(\hat\bma_n-\bma_0\right),
\end{align*}
Likewise,
\begin{align*}
&\int\left\{\bmx-\E\left(\bm X|\hat\bma_n^T\bm x\right)\right\}\left\{y-\hat\psi_{n\hat\bma_n}(\hat\bma_n^T\bmx)\right\}
\,d\bigl(\P_n-P_0\bigr)(\bmx,y)\\
&=\int\left\{\bmx-\E\left(\bm X|\bma_0^T\bm x\right)\right\}\left\{y-\hat\psi_{n\hat\bma_n}(\hat\bma_n^T\bmx)\right\}
\,d\bigl(\P_n-P_0\bigr)(\bmx,y)+o_p\left(\hat\bma_n-\bma_0\right)\\
&=\int\left\{\bmx-\E\left(\bm X|\bma_0^T\bm x\right)\right\}\left\{y-\psi_0(\bma_0^T\bmx)\right\}
\,d\bigl(\P_n-P_0\bigr)(\bmx,y)+o_p\left(\hat\bma_n-\bma_0\right)\\
&\qquad+\int\left\{\bmx-\E\left(\bm X|\bma_0^T\bm x\right)\right\}\left\{\psi_0(\bma_0^T\bmx)-\hat\psi_{n\hat\bma_n}(\hat\bma_n^T\bmx)\right\}
\,d\bigl(\P_n-P_0\bigr)(\bmx,y)+o_p\left(\hat\bma_n-\bma_0\right),
\end{align*}
where the last dominating term satisfies
\begin{align*}
&\int\left\{\bmx-\E\left(\bm X|\bma_0^T\bm x\right)\right\}\left\{\psi_0(\bma_0^T\bmx)-\hat\psi_{n\hat\bma_n}(\hat\bma_n^T\bmx)\right\}
\,d\bigl(\P_n-P_0\bigr)(\bmx,y)\\
&=\int\left\{\bmx-\E\left(\bm X|\bma_0^T\bm x\right)\right\}\left\{\psi_0(\bma_0^T\bmx)-\psi_{\hat\bma_n}(\hat\bma_n^T\bmx)\right\}
\,d\bigl(\P_n-P_0\bigr)(\bmx,y)\\
&\qquad+\int\left\{\bmx-\E\left(\bm X|\bma_0^T\bm x\right)\right\}\left\{\psi_{\hat\bma_n}(\hat\bma_n^T\bmx)-\hat\psi_{n\hat\bma_n}(\hat\bma_n^T\bmx)\right\}
\,d\bigl(\P_n-P_0\bigr)(\bmx,y)\\
&=o_p\left(\hat\bma_n-\bma_0\right)+o_p\left(n^{-1/2}\right).
\end{align*}
So we arrive at the following asymptotic equation:
\begin{align}
\label{SSE_equation}
&\E\left\{\psi_0'(\bma_0^T\bmX)\left\{\bmX-\E\left(\bm X|\bma_0^T\bm x\right)\right\}\left\{\bmX-\E\left(\bm X|\bma_0^T\bm X\right)\right\}^T\right\}\left\{\hat\bma_n-\bma_0\right\}\nonumber\\
&=\int\left\{\bmx-\E\left(\bm X|\bma_0^T\bm x\right)\right\}\left\{y-\psi_0(\bma_0^T\bmx)\right\}
\,d\bigl(\P_n-P_0\bigr)(\bmx,y)
+o_p\left(\hat\bma_n-\bma_0\right)+o_p\left(n^{-1/2}\right).
\end{align}

Defining
\begin{align*}
&\bm B=\E\left\{\bmX-\E\left(\bm X|\bma_0^T\bm X\right)\right\}\left\{\bmX-\E\left(\bm X|\bma_0^T\bm X\right)\right\}^T\psi_0'(\bma_0^T\bmX),
\end{align*}
we get:
\begin{align*}
\hat\bma_n-\bma_0=k\bma_0-\bm{B}^-\int\left\{\bmx-\E\left(\bm X|\bma_0^T\bm x\right)\right\}\left\{y-\psi_0(\bma_0^T\bmx)\right\}
\,d\bigl(\P_n-P_0\bigr)(\bmx,y)+o_p\left(n^{-1/2}\right)+o_p\left(\hat\bma_n-\bma_0\right),
\end{align*}
for some $k\in\R$, where $\bmB^-$ is the Moore-Penrose inverse of $\bmB$.

Define
\begin{align*}
\bmb_n=-\bm{B}^-\int\left\{\bmx-\E\left(\bm X|\bma_0^T\bm x\right)\right\}\left\{y-\psi_0(\bma_0^T\bmx)\right\}
\,d\bigl(\P_n-P_0\bigr)(\bmx,y).
\end{align*}
Then, since $\bmb_n$ is perpendicular to $\bma_0$, we get:
\begin{align*}
\|\hat\bma_n\|_2^2=1=(1+k)^2\|\bma_0\|_2^2+\|\bmb_n\|_2^2=(1+k)^2+\|\bmb_n\|_2^2,
\end{align*}
implying
\begin{align*}
k=-1\pm\sqrt{1-\|\bmb_n\|_2^2}.
\end{align*}
and hence
\begin{align*}
k=-\tfrac12\|\bmb_n\|_2^2\left\{1+o_p(1)\right\}\qquad\text{or}\qquad k=-2+\tfrac12\|\bmb_n\|_2^2\left\{1+o_p(1)\right\}.
\end{align*}
The second solution would lead to
\begin{align*}
\hat\bma_n&=-\bma_0+\tfrac12\|\bmb_n\|_2^2\left\{1+o_p(1)\right\}+O_p\left(n^{-1/2}\right)+o_p\left(\hat\bma_n-\bma_0\right)\\
&=-\bma_0+O_p\left(n^{-1/2}\right)+o_p\left(\hat\bma_n-\bma_0\right)
\end{align*}
which is excluded by $\hat\bma_n-\bma_0=o_p(1)$ (the consistency of $\hat\bma_n$). Using $\|\bmb_n\|_2^2=O_p(n^{-1})$ we get:
\begin{align*}
\hat\bma_n-\bma_0&=-\tfrac12\|\bmb_n\|_2^2\bma_0+\bmb_n+o_p\left(n^{-1/2}\right)+o_p\left(\hat\bma_n-\bma_0\right)\\
&=\bmb_n+o_p\left(n^{-1/2}\right)+o_p\left(\hat\bma_n-\bma_0\right),
\end{align*}
and hence
\begin{align*}
\hat\bma_n-\bma_0&=-\bm{B}^-\int\left\{\bmx-\E\left(\bm X|\bma_0^T\bm x\right)\right\}\left\{y-\psi_0(\bma_0^T\bmx)\right\}
\,d\bigl(\P_n-P_0\bigr)(\bmx,y)\\
&\qquad\qquad+o_p\left(n^{-1/2}\right)+o_p\left(\hat\bma_n-\bma_0\right).
\end{align*}
\end{proof}

\section{Consistency and asymptotic normality of the Efficient Score Estimate (ESE) and the Penalized Least Squares Estimate (PLSE) as defined by (\ref{spline_eq})}
\label{sec:ESE_PLSE}
The proofs of the consistency and asymptotic normality of the ESE and PLSE are highly similar to the proofs of these facts for the SSE in the preceding section. The only extra ingredient is occurrence of the estimate of the derivative of the link function. We only discuss the asymptotic normality.

An essential step is again to show that
\begin{align*}
&\int\bmx\left\{y-\hat\psi_{n,\hat\bma_n}(\hat\bma_n^T\bmx)\right\}\hat\psi'_{n\hat\bma_n}(\hat\bma_n^T\bmx)\,d\P_n(\bmx,y)\\
&=\int\left\{\bmx-\E(X|\hat\bma_n^T\bmX)\right\}\left\{y-\hat\psi_{n,\hat\bma_n}(\hat\bma_n^T\bmx)\right\}\hat\psi'_{n\hat\bma_n}(\hat\bma_n^T\bmx)\,d\P_n(\bmx,y)
+o_p(n^{-1/2})+o_p(\hat\bma_n-\bma_0),
\end{align*}
For the ESE this is done by defining the piecewise constant function $\bar{\rho}_{n, \bma}$ for $u$ in the interval between successive jumps $\tau_i$  and $\tau_{i+1})$  of $\hat\psi_{n\bma}$ by:
\begin{eqnarray*}
	\bar {\rho}_{n, \bma}(u)  =  \left \{
	\begin{array}{lll}
		\E[\bm X| \bma^T\bm X= \t_i]\psi_{\bma}'(\t_i) \ \  \ \ \ \ \ \ \textrm{ if $\psi_{\bma}(u)  > \hat\psi_{n\bma}(\tau_i)$  \ for all $u \in (\tau_i, \tau_{i+1})$}, \\
		\E[\bm X| \bma^T\bm X= s]\psi_{\bma}'(s) \ \ \ \  \ \  \  \ \ \ \textrm{ if $\psi_{\bma}(s)  = \hat\psi_{n\bma}(s)$ \ for some $s \in (\tau_i, \tau_{i+1})$}, \\
		\E[\bm X| \bma^T\bm X= \t_{i+1}]\psi_{\bma}'(\t_{i+1})\ \ \ \textrm{if $\psi_{\bma}(u) < \hat\psi_{n\bma}(\tau_i)$  \ for all $u \in (\tau_i, \tau_{i+1})$}. 
	\end{array}
	\right.
\end{eqnarray*}
where $\bar{\rho}_{n,\bma}$ replaces $\bar E_{n, \bma}$ in (\ref{E_n-def}), see Appendix E in the supplement of \cite{balabdaoui2018}. The remaining part of the proof runs along the same lines as the proof for the SSE. For additional details, see Appendix E in the supplement of \cite{balabdaoui2018}.

The corresponding step in the proof for the PLSE is given by the following lemma.

\begin{lemma}
Let the conditions of Theorem 5 in  \cite{Kuchibhotla_patra:17} be satisfied. In particular, let the penalty parameter $\m_n$ satisfy $\m_n=o_p(n^{-1/2})$. Then we have for all $\bma$ in a neighborhood of $\bma_0$ and for the corresponding penalized least squares estimate (natural cubic spline)  $\hat\psi_{n\bma}$:
\begin{align*}
\int\E(\bmX|\bma^T\bmX)\left\{y-\hat\psi_{n\bma}\left(\bma^T\bmx\right)\right\}\hat\psi_{n\bma}'\left(\bma^T\bmx\right)\,d\P_n(\bmx,y)=O_p(\mu_n)=o_p\left(n^{-1/2}\right).
\end{align*}
\end{lemma}

\begin{remark}
{\rm The result shows that we have as our  basic equation in $\bma$:
\begin{align*}
&\frac1n\left(\bm I-\bma\bma^T\right)\sum_{i=1}^n \bigl\{\hat\psi_{n\bma}(\bma^T\bmX_i)-Y_i\bigr\}\hat\psi'_{n\bma}(\bma^T\bmX_i)\bmX_i\\
&=\frac1n\left(\bm I-\bma\bma^T\right)\sum_{i=1}^n \bigl\{\hat\psi_{n\bma}(\bma^T\bmX_i)-Y_i\bigr\}\hat\psi'_{n\bma}(\bma^T\bmX_i)
\left\{\bmX_i-\E(\bmX_i|\bma^T\bmX_i)\right\}+o_p\left(n^{-1/2}\right)\\
&=o_p\left(n^{-1/2}\right).
\end{align*}
}
\end{remark}

\begin{proof}
Fix $\bma$ and let $\hat\psi_{n,\bma}$ be the natural cubic spline, minimizing
\begin{align*}
n^{-1}\sum_{i=1}^n\left\{Y_i-f(t_i)\right\}^2+\mu_n\int_a^b \left\{f''(t)\right\}^2\,dt,
\end{align*}
over functions $g\in{\cal S}_2[a,b]$, where the $t_i$ are the the ordered values $\bma^T\bmX_i$, and where $\m_n=o_p(n^{-1/2})$, and where $a=\min_i\bma^T\bmX_i$ and $b=\max_i\bma^T\bmX_i$. We can write the minimum in the following form:
\begin{align*}
&\int\left\{y-\hat\psi_{n\bma}(\bma^T\bmx)\right\}^2\,d\P_n(\bmx,y)+\mu_n\int_a^b \left\{\hat\psi_{n\bma}''(t)\right\}^2\,dt,
\end{align*}
We extend the natural cubic spline $\hat\psi_{n,\bma}$ linearly to a function on $\R$, and define the function
\begin{align}
\label{spline_derivative}
\bm v\mapsto \f(\bm v)=&\int\left\{y-\hat\psi_{n\bma}\left(\bma^T\bmx+\bm v^T\,\E(\bmX|\bma^T\bmX)\right)\right\}^2\,d\P_n(\bmx,y)\\\
&\qquad+\mu_n\int_a^b \left\{\hat\psi_{n\bma}''\left(t+\bm v^T\,\E(\bmX|\bma^T\bmX=t)\right)\right\}^2\,dt,
\end{align}
We have:
\begin{align*}
&\frac{\partial}{\partial\bmv} \int_a^b \left\{\hat\psi_{n\bma}''\left(t+\bm v^T\,\E(\bmX|\bma^T\bmX=t)\right)\right\}^2\,dt\biggr|_{\bmv=\bm0}\\
&=2\int_a^b \hat\psi_{n\bma}'''(t)\hat\psi_{n\bma}''(t)\E(\bmX|\bma^T\bmX=t)\,dt
=-\int_a^b \hat\psi_{n\bma}''(t)^2\frac{\partial}{\partial t}\E(\bmX|\bma^T\bmX=t)\,dt.
\end{align*}
This implies, assuming the boundedness of the derivative of the function $t\mapsto\frac{\partial}{\partial t}\E(\bmX|\bma^T\bmX=t)$ for $t\in[a,b]$,
\begin{align*}
&\left|\frac{\partial}{\partial\bmv} \int_a^b \left\{\hat\psi_{n\bma}''\left(t+\bm v^T\,\E(\bmX|\bma^T\bmX=t)\right)\right\}^2\,dt\biggr|_{\bmv=\bm0}\right|
\lesssim\int_a^b \left\{\hat\psi_{n\bma}''(t)\right\}^2\,dt.
\end{align*}
Also assuming that
\begin{align*}
\int_a^b \left\{\hat\psi_{n\bma}''(t)\right\}^2\,dt=O_p(1),
\end{align*}
(see Theorem 2 in \cite{Kuchibhotla_patra:17}), we obtain from (\ref{spline_derivative}):
\begin{align*}
\f'(\bm0)=\int\E(\bmX|\bma^T\bmX)\left\{y-\hat\psi_{n\bma}\left(\bma^T\bmx\right)\right\}\hat\psi_{n\bma}'\left(\bma^T\bmx\right)\,d\P_n(\bmx,y)+O_p\left(\mu_n\right)=\bm0,
\end{align*}
since the function $\f$ attains its minimum at $\bm0$ by the definition of the (natural) cubic spline as a least squares estimate. It follows that
\begin{align}
\label{spline_conditional_expectation}
\int\E(\bmX|\bma^T\bmX)\left\{y-\hat\psi_{n\bma}\left(\bma^T\bmx\right)\right\}\hat\psi'_{n\bma}\left(\bma^T\bmx\right)\,d\P_n(\bmx,y)=
O_p\left(\mu_n\right)=o_p\left(n^{-1/2}\right).
\end{align}
\end{proof}

The remaining part of the proof of the asymptotic normality can either run along the same lines as the proof for the corresponding fact for the SSE, using the function $u\mapsto\psi_{\bma}(u)=\E\{\psi_0(\bma^T\bmx)|\bma^T\bmX=u\}$, or directly use the convergence of $\hat\psi_{n\hat\bma_n}$ to $\psi_0$ and of $\hat\psi'_{n\hat\bma_n}$ to $\psi_0'$ (see Theorem 3 in \cite{Kuchibhotla_patra:17}). For the SSE and ESE we were forced to introduce the intermediate function $\psi_{\bma}$ to get to the derivatives, because there the derivative of $\hat\psi_{n\hat\bma_n}$ did not exist.

We get the following result.

\begin{theorem}
	\label{theorem:asymptotics-efficient}
	Let either $\hat\bma_n$ be the ESE of $\bma_0$ and  let Assumptions A1-A8 of \cite{balabdaoui2018} be satisfied and let $\psi_{\bma}$ be twice continuously differentiable or let $\hat\bma_n$ be the PLSE of $\bma_0$ and let Assumptions (A0) to (A6) and (B1) to (B3) of  \cite{Kuchibhotla_patra:17}) be satisfied.
	Moreover, let the bandwidth $h \asymp n^{-1/7}$ in the estimate of the derivative of $\psi_{\bma}$ for the ESE.
			Define the matrices,
		\begin{align}
		\label{def:I1}
		\tilde {\bm A}:=\E\Bigl[\psi_0'(\bm\a_0^T\bm X)^2\,\text{\rm Cov}(\bm X|\bm\a_0^T\bm X)\Bigr],
		\end{align}
		and
		\begin{align}
		\label{def:I2}
		\tilde {\bm \Sigma}:=\E\left[\left\{Y -\psi_0(\bm\a_0^T\bm X)\right\}^2\psi_0'(\bm\a_0^T\bm X)^2\left\{\bm X -\E(\bm X|\bm\a_0^T\bm X) \right\}\left\{\bm X -\E(\bm X|\bm\a_0^T\bm X) \right\}^T\right].
		\end{align}
		Then 
		\begin{align*}
		\sqrt n (\tilde \bma_n - \bma_0) \to_d N_{d}\left(\bm 0,  \tilde {\bm A}^- \tilde {\bm \Sigma} \tilde {\bm A}^-\right),
		\end{align*}
		where $\tilde {\bm A}^{-}$ is the Moore-Penrose inverse of $\tilde {\bm A}$.	
\end{theorem}

This corresponds to Theorem 6 in \cite{balabdaoui2018} and Theorem 5 in \cite{Kuchibhotla_patra:17}), but note that the formulation of Theorem 5 in \cite{Kuchibhotla_patra:17} still contains  the Jacobian connected with the lower dimensional parametrization.

\section{Simulations}
\label{section:simulations}
In this section we give results similar to those in Section 7 of \cite{balabdaoui2018}. First we want to show that the Lagrange approach and the random starting values give results similar to those in Table 2 of \cite{balabdaoui2018} for the model introduced in Section \ref{sec:example} of the present paper. This can be seen from Table \ref{table:simulation1}, where the results for the SSE, ESE, LSE, MRE and linear estimate can be compared with the corresponding results inTable 2 of  \cite{balabdaoui2018}. The MRE is called MRCE   (Maximum Rank Correlation Estimator) and the linear estimator is called H-LFLSE (``Hybrid Link-Free Least Squares Estimator'') in Table 2 of \cite{balabdaoui2018}. Moreover, we have added the results for the penalized least squares estimator, as computed via the algorithm in \cite{github:18} (PLSE) and via the {\tt R} package {\tt simest} (PLSE2), respectively.

\begin{table}[!ht]
	\centering
	\caption{ Simulation, model  ($X_i\sim  N(0,1), d=3$): The mean value ($\hat\mu_i$ =  mean($\hat \a_{in}), i=1,2,3$) and $n$ times the variance-covariance ($\hat\sigma_{ij} = n\cdot$cov$(\hat\a_{in},\hat\a_{jn})$,$i,j=1,2,3$) of the simple score estimate (SSE), the efficient score estimate (ESE), the least squares estimate (LSE), the maximum rank correlation estimate (MRE), the linear estimate (this is H-LFLSE in \cite{balabdaoui2018}), the Penalized Least Squares Estimate (PLSE),  the Penalized Least Squares Estimate as computed by the {\tt R} package {\tt simest} (PLSE2) and the Efficient Dimension Reduction estimate as computed by the {\tt R} package {\tt edr} (EDR), for different sample sizes $n$ with  $X_i\sim  N(0,1)$. The line, preceded by $\infty$, gives the asymptotic values. The values are based on $1000$ replications.}
	\label{table:simulation1}
	\vspace{0.5cm}
	\scalebox{1}{
		\begin{tabular}{|lr|ccc|cccccc |}
			\hline
			Method&$n$ & $\hat\mu_1$ & $\hat\mu_2$ & $\hat\mu_3$ & $\hat \sigma_{11}$& $\hat\sigma_{22}$& $\hat\sigma_{33}$& $\hat\sigma_{12}$&$\hat\sigma_{13}$&$\hat\sigma_{23}$\\
			\hline
			&&&&&&&&&&\\
			SSE&100 & 0.5735 & 0.5769 & 0.5752 &0.2434& 0.2482& 0.2399 & -0.1234 & -0.1165& -0.1229 \\
			&500 & 0.5771 & 0.5768 & 0.5774 & 0.1493 & 0.1426 & 0.1470 & -0.0721 & -0.0771 & -0.0700 \\
			&1000 & 0.5772 & 0.5776 & 0.5770 & 0.1230 & 0.1150 & 0.1268 & -0.0558 & -0.0671 & -0.0594 \\
			&2000& 0.5770 & 0.5775 & 0.5775 & 0.1118 & 0.1091 & 0.1004 & -0.0600 & -0.0516 & -0.0491\\
			\hline
			&$\infty$&0.5774 & 0.5774 & 0.5774 & 0.0741 & 0.0741 & 0.0741 & -0.0370 & -0.0370 & -0.0370\\
			&&&&&&&&&&\\
			&&&&&&&&&&\\
			ESE&100 & 0.5756 & 0.5779 & 0.5758 & 0.1061 & 0.1058 & 0.1105 & -0.0510 &-0.0549 & -0.0548\\
			&500 & 0.5773 & 0.5770 & 0.5776 & 0.0491 & 0.0489 & 0.0518 & -0.0231 & -0.0261 & -0.0258 \\
			&1000 & 0.5775 & 0.5773 & 0.5771 & 0.0438 & 0.0406 & 0.0420 & -0.0212 & -0.0226 & -0.0193 \\
			&2000& 0.5773 & 0.5773 & 0.5774 & 0.0362 & 0.0367 & 0.0327 & -0.0201 & -0.0161 & -0.0166\\
			\hline
			&$\infty$&0.5774 & 0.5774 & 0.5774 & 0.0247 & 0.0247 & 0.0247 & -0.0123 & -0.0123 & -0.0123\\
			&&&&&&&&&&\\
			&&&&&&&&&&\\
			LSE&100 & 0.5751 & 0.5766 & 0.5759 & 0.1627 & 0.1729 & 0.1753 &-0.0799& -0.0820& -0.0923\\
			&500 & 0.5773 & 0.5770 & 0.5771 & 0.1048 & 0.1100 & 0.1044 & -0.0550 & -0.0497 & -0.0548 \\
			&1000 & 0.5773 & 0.5776 & 0.5761 & 0.0985 & 0.0884 & 0.0995 & -0.0437 & -0.0550 & -0.0445 \\
			&2000& 0.5775 & 0.5772 & 0.5772 & 0.0921 & 0.0968 & 0.0930 & -0.0474& -0.0436 &-0.0494\\
			\hline
			&$\infty$&0.5774 & 0.5774 & 0.5774 & ? & ?& ? & ? & ? &? \\
			&&&&&&&&&&\\
			&&&&&&&&&&\\
			MRE&100 &0.5683&  0.5731 & 0.5736 & 0.6614 & 0.6764&  0.6181& -0.3258 &-0.2999& -0.3235\\
			&500 & 0.5760 & 0.5770 & 0.5763 & 0.5208 & 0.4969 & 0.5125 & -0.2521 & -0.2670 & -0.2437 \\
			&1000 & 0.5771 & 0.5776 & 0.5761 & 0.4815 & 0.4512 & 0.4842 & -0.2259 & -0.2555 & -0.2262 \\
			&2000 &0.5772 & 0.5770  &0.5773 & 0.4704  &0.4502 &0.4170 &-0.2506 &-0.2186&-0.1996\\
			\hline
			&$\infty$&0.5774 & 0.5774 & 0.5774 & 0.3583 & 0.3583& 0.3583 & -0.1791 & -0.1791 & -0.1791\\
			&&&&&&&&&&\\
			&&&&&&&&&&\\
			linear & 100 & 0.5746 & 0.5736 & 0.5724 & 0.4675 & 0.4588 & 0.4644 &-0.2309&-0.2323& -0.2243 \\
			&500 & 0.5780 & 0.5751 & 0.5762 & 0.5072 & 0.5057 & 0.4966 & -0.2635 & -0.2733 & -0.2445 \\
			&1000 & 0.5771 & 0.5768 & 0.5768 & 0.5182 & 0.5099 & 0.5085 & -0.2600 & -0.2583 & -0.2491 \\
			& 2000 & 0.5772&  0.5770 & 0.5710 & 0.5129 & 0.5288 & 0.5067 &-0.2673 &-0.2453& -0.2769 \\
			\hline
			&$\infty$&0.5774 & 0.5774 & 0.5774 & 0.5185 & 0.5185 & 0.5185 & -0.2593 & -0.2593 & -0.2593\\
		&&&&&&&&&&\\
		&&&&&&&&&&\\			
		PLSE & 100&0.5765&0.5772&0.5766&0.0657&0.0682&0.0629&-0.0352&-0.0303&-0.0326\\
		&500 & 0.5775 & 0.5770 & 0.5774 & 0.0368 & 0.0362 & 0.0384 & -0.0173 & -0.0195 & -0.0189 \\
		&1000 & 0.5775 & 0.5774 & 0.5771 & 0.0329 & 0.0314 & 0.0305 & -0.0169 & -0.0160& -0.0145 \\
		& 2000 & 0.5773&  0.5775 & 0.5774 & 0.0301 & 0.0313 & 0.0282 &-0.0166 &-0.0135& -0.0147 \\
		\hline
		&$\infty$&0.5774 & 0.5774 & 0.5773 & 0.0276 & 0.0278 & 0.0282 & -0.0137 & -0.0140 & -0.0142\\
		&&&&&&&&&&\\
			&&&&&&&&&&\\
			PLSE2 & 100 &0.5767& 0.5771  & 0.5766 & 0.0642 &0.0661&0.0614 &--0.0342& -0.0301 &-0.0314\\
			&500 & 0.5777 & 0.5768 & 0.5773 & 0.0496 & 0.0485 & 0.0505 & -0.2635 & -0.2733 & -0.2445 \\
			&1000 & 0.5776 & 0.5774 & 0.5771 & 0.0609 & 0.0542 & 0.0565 & -0.0169 & -0.0317 & -0.0248 \\
			& 2000 & 0.5771&  0.5777 & 0.5772 & 0.0724 & 0.0713 & 0.0705 &-0.0365 &-0.0356& -0.0350 \\
			\hline
			&$\infty$&0.5774 & 0.5774 & 0.5774 & 0.0247 & 0.0247 & 0.0247 & -0.0123 & -0.0123 & -0.0123\\
		&&&&&&&&&&\\
		&&&&&&&&&&\\
			EDR&100 &0.5735&  0.5755 & 0.5783 &0.1771 & 0.1827  &0.1836& -0.0815 &-0.0882&  -0.0993\\
			&500 & 0.5772 & 0.5770 & 0.5775 & 0.0609 & 0.0598 & 0.0642 & -0.0293 & -0.0328 & -0.0315 \\
			&1000 & 0.5773 & 0.5776 & 0.5769 & 0.0515 & 0.0486 & 0.0521 & -0.0241 & -0.0275 & -0.0245 \\
			&2000 &0.5773 & 0.5774  &0.5773 & 0.0450  &0.0439 & 0.0417 &-0.0236 &-0.0214 &-0.0417\\
			\hline
			&$\infty$ &0.5774 & 0.5774 & 0.5774 & ? & ?& ? & ? & ? & ?\\
			&&&&&&&&&&\\
		\hline
		\end{tabular}}
	\end{table}

It is clear that there is good correspondence between the results. It is also of interest to see (again) that the linear estimator and Han's maximumum rank correlation estimator have a behavior that is inferior to that of the other estimators. The good behavior of the penalized least squares estimators is also of interest. The penalty parameter $\mu$ was taken $0.1$ for sample sizes $100, 500$ and $1000$ and taken $0.05$ for $n=2000$, but the results do not seem very sensitive to this choice. For exampe, if we take $\mu=0.1$ also for sample size $n=2000$, the results are almost the same. It is also seen that the behavior of the PLSE as computed via the algorithm in \cite{github:18} has a slightly better behavior for the larger sample sizes, possibly because the solution via equation (\ref{spline_eq}), motivated by the Lagrange-type approach, gives a more stable behavior than the application of the argmax approach for $\bma$. The splines $\hat\psi_{n,\bma}$ are the same in the two methods, but the search for the best $\bma$ is  different, also because different search methods are used; \cite{github:18} uses the Hooke-Jeeves algorithm and {\tt simest} uses Broyden's algorithm.

\begin{figure}[!h]
	\centering
	\begin{subfigure}{0.485\linewidth}
		\includegraphics[width=0.95\textwidth]{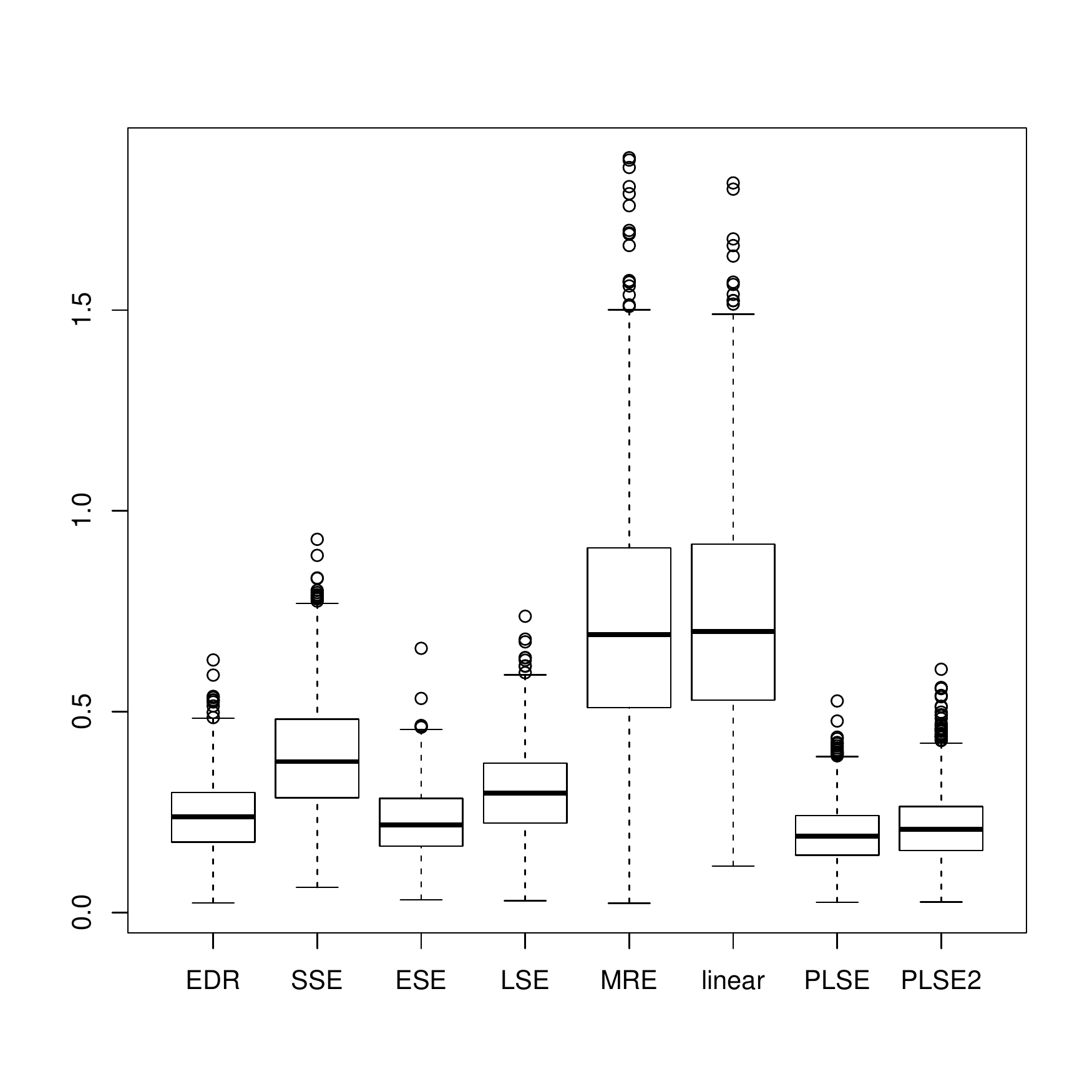}
		\caption{$d = 5$}
	\end{subfigure}
	\begin{subfigure}{0.485\linewidth}
	\includegraphics[width=0.95\textwidth]{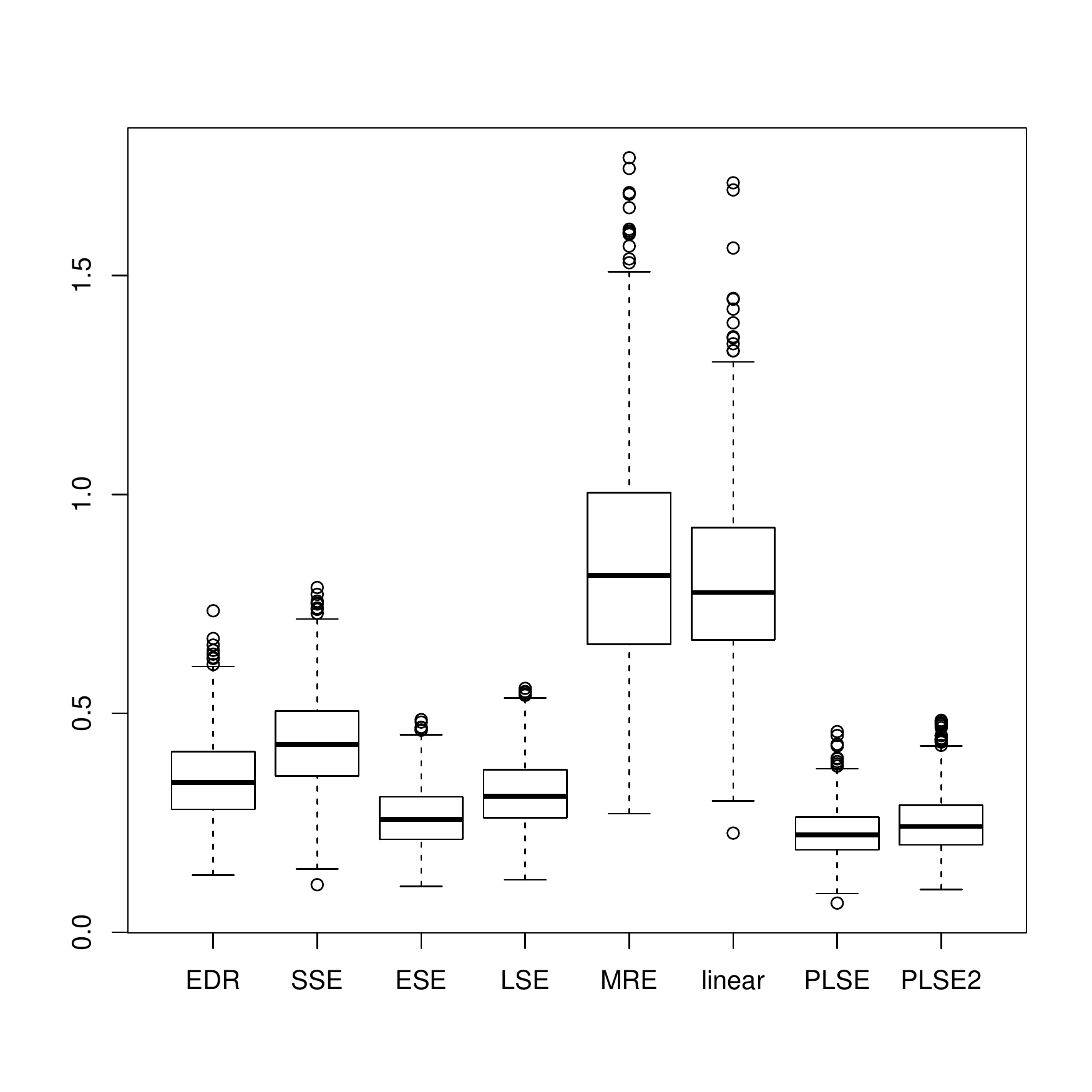}
	\caption{$d = 10$}
\end{subfigure}\\
	\begin{subfigure}{0.485\linewidth}
		\includegraphics[width=0.95\textwidth]{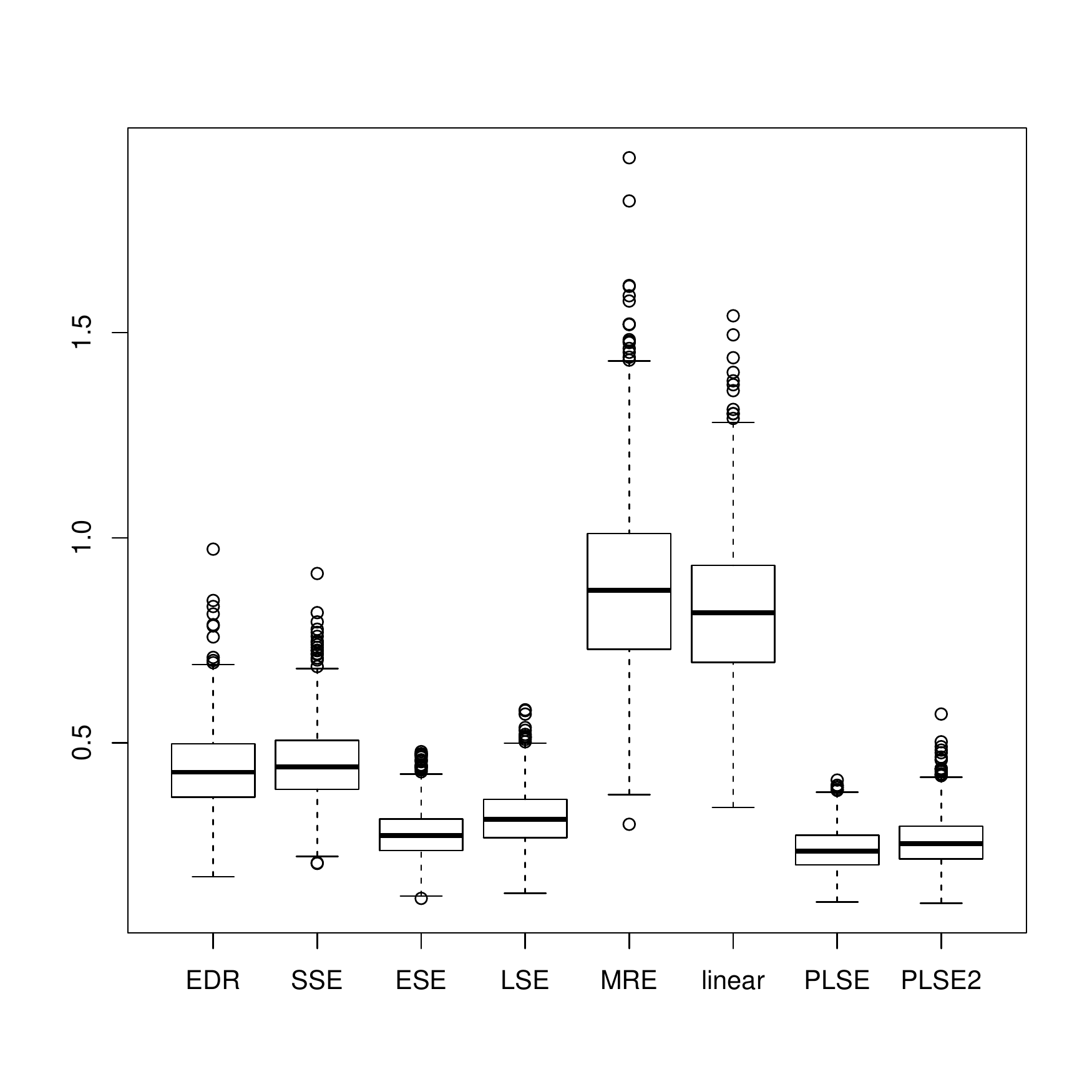}
		\caption{$d=15$}
	\end{subfigure}
\begin{subfigure}{0.485\linewidth}
	\includegraphics[width=0.95\textwidth]{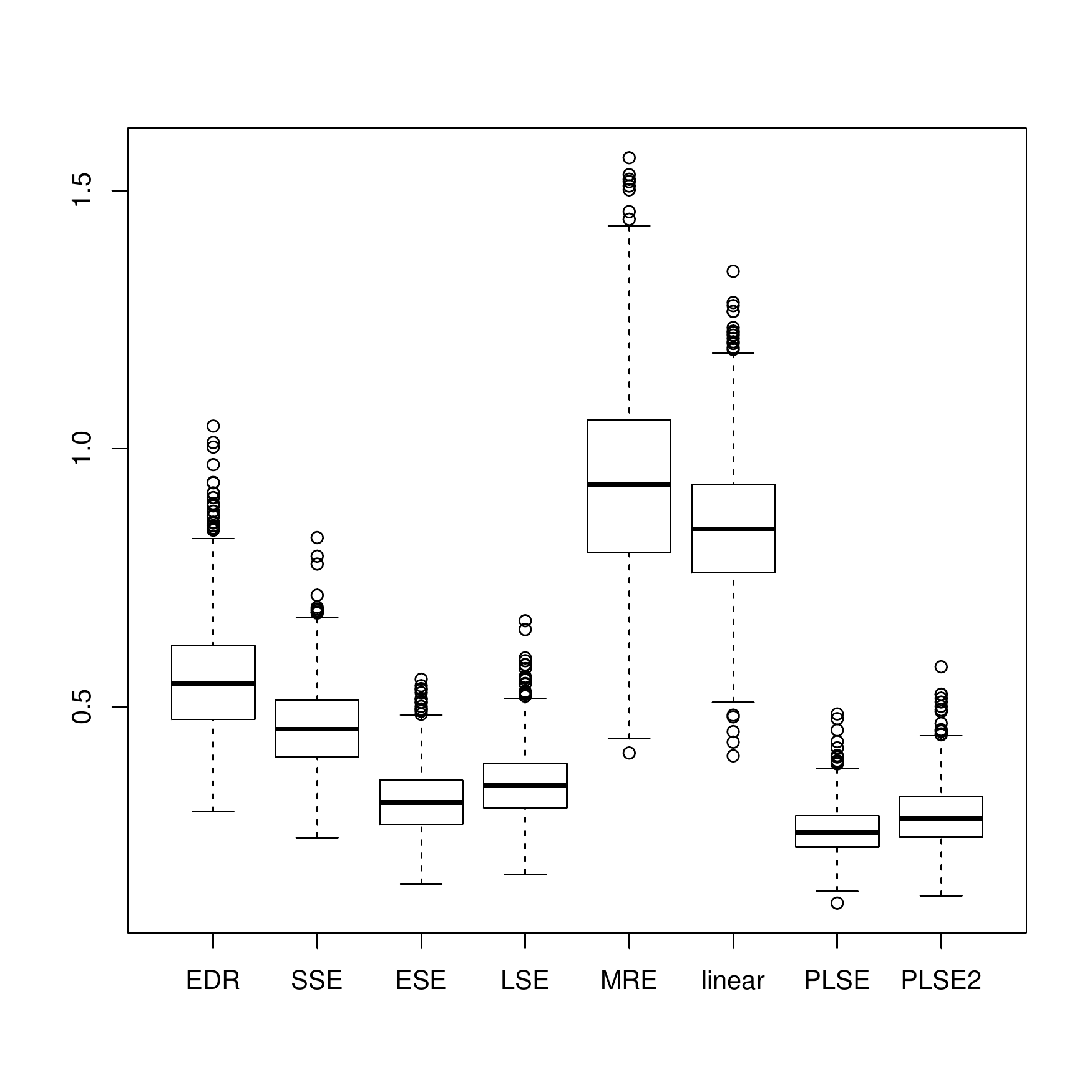}
	\caption{$d = 25$}
\end{subfigure}
	\caption{Boxplots of $\sqrt{n/d}\,\|\hat\bma_n-\bma_0\|_2$ for $n=500$ and for (a) $d = 5$, (b) $d=10$, (c) $d=15$ and (d) $d=25$ and $1000$ replications for the EDR, SSE, ESE, LSE, MRE, the linear estimate, the PLSE and the PLSE, as computed by the {\tt R} package {\tt simest} (PLSE2).
	All algorithms except EDR and LSE were started at the value given by the LSE. The estimate $\hat\bma_n$ of the LSE was chosen as the value giving the smallest least squares criterion in the computations from $20$ random starting values.}
\label{boxplots2}
\end{figure}

Just as in \cite{balabdaoui2018}, Section 7.2, we investigate the effect of an increasing dimension for this model. The result is given in Figure \ref{boxplots2}. The inferiority of the linear estimator and of Han's maximum rank correlation estimator is again clear. The estimator EDR is deteriorating in comparison with the estimators SSE, ESE, LSE, PLSE and PLSE2 with increasing dimension and, moreover, its computing time increases steeply with increasing dimension (see also for this Section 7.2 of \cite{balabdaoui2018}).

There is again some evidence that the PLSE, computed by the algorithm in \cite{github:18} has a slightly better behavior than the PLSE2, computed by {\tt simest}. Both estimates seem to have the best behavior among the estimates studied here, though, with the ESE as a close second.

\section{Conclusion}
\label{section:conclusion}
A natural estimate  the finite dimensional parameter in the monotone single index is the profile least squares estimator which we called the LSE. But its asymptotic distribution is unknown and  it is even not known whether this estimate is $\sqrt{n}$ consistent. The LSE is an $M$ estimate in two senses: for fixed $\bma$ the least squares criterion (\ref{step1})
is minimized over all monotone functions $\psi$ and the function (\ref{step2}) is minimized for the resulting $\hat\psi_{n,\bma}$  over $\bma$. 

We proposed to replace the second minimization by the solution of an equation arising from a Lagrange formulation of the minimization problem. This method yields estimators for which one can prove $\sqrt{n}$ convergence and asymptotic normality. Introducing the estimate of the derivative of the link function (\ref{estimate_derivative_psi}), one can even construct an asymptotically efficient estimate in this way. The method avoids reparametrization, which is the usual way to go in this model, and continues work in \cite{balabdaoui2018}.

It is shown that one can also apply this approach for a method which uses cubic splines in the first minimization step, which then becomes a penalized minimization problem. Our method avoids again reparametrization and continues work in this direction in \cite{Kuchibhotla_patra:17}, which was implemented in the {\tt R} package {\tt simest}.

The methods are illustrated by simulations, and the implementation  can be found on \cite{github:18} in the form of {\tt R} scripts and {\tt C++} code. We also address the important issue of the starting points of the algorithms and show that one can start  by using the result of the LSE as starting value. The LSE itself is computed from a number of random starting values and the estimate resulting from the starting value which gives a minimum of the criterion function is used as starting value for the other algorithms.

The estimates are compared with a number of estimates in the literature: the estimate  in the {\tt R} package {\tt simest}, Han's maximum rank correlation estimate (see \cite{han:87}), a linear estimate, applicable if the distribution of the covariate is elliptically symmetric, and the effective dimension reduction estimate, introduced in \cite{hristache01}, and available via the {\tt R} package {\tt edr}. In our comparison the penalized least squares estimates come out best, with the efficient score estimate ESE as a close second. The linear estimate and Han's maximum rank correlation estimate seem clearly inferior to the other estimators.

In the simulation examples we considered covariates with a normal distribution, as a particular example of elliptically symmetric distributions, since in that case the conditions for the application of the linear method are satisfied. We also took normal errors, since this was a condition for the application of {\tt edr} with which we also make a comparison. Other comparisons with {\tt edr} and other packages can be found in \cite{Kuchibhotla_patra:17} and \cite{balabdaoui2018}. A general finding is that the linear method completely breaks down if the condition of elliptic symmetry of the covariates is not fulfillled.

Further improvement of our Lagrange-type method is to be expected if we refine the search algorithms. Although most of the programming was done from first principles, in particular the programming of the spline method which used the Reinsch method, as exposed in \cite{green_silverman:94}, to make effective use of the band matrix structure, the Hooke-Jeeves and Nelder-Mead {\tt C++} algorithms were ``taken off the (internet) shelf'' (see \cite{github:18}), and certainly can be improved upon. Because the criterion function is not continuous, at least for the SSE, ESE and LSE, doing the search for $\bma$ in the right way is rather non-trivial. Changes of $\bma$ lead to changes of the order of the values $\bma^T\bmX_i$ and this leads to discontinuities of the criterion function.  Note that these searches are ``grid-free'', looking for a solution on a grid is a much too crude method in these problems.

\bibliographystyle{imsart-nameyear}
\bibliography{cupbook}

\begin{thebibliography}{9}

\bibitem[\protect\citeauthoryear{Balabdaoui, Durot and
  Jankowski}{2016}]{Balabdaoui_Durot_Jankowski:2016}
\begin{barticle}[author]
\bauthor{\bsnm{Balabdaoui},~\bfnm{Fadoua}\binits{F.}},
  \bauthor{\bsnm{Durot},~\bfnm{C{\'e}cile}\binits{C.}} \AND
  \bauthor{\bsnm{Jankowski},~\bfnm{Hanna}\binits{H.}}
(\byear{2016}).
\btitle{Least squares estimation in the monotone single index model}.
\bjournal{arXiv preprint arXiv:1610.06026}.
\end{barticle}
\endbibitem

\bibitem[\protect\citeauthoryear{Balabdaoui, Groeneboom and
  Hendrickx}{2018}]{balabdaoui2018}
\begin{bmisc}[author]
\bauthor{\bsnm{Balabdaoui},~\bfnm{Fadoua}\binits{F.}},
  \bauthor{\bsnm{Groeneboom},~\bfnm{Piet}\binits{P.}} \AND
  \bauthor{\bsnm{Hendrickx},~\bfnm{Kim}\binits{K.}}
(\byear{2018}).
\btitle{Score estimation in the monotone single index model}.
\bhowpublished{https://doi.org/10.1111/sjos.12361}.
\bdoi{10.1111/sjos.12361}
\end{bmisc}
\endbibitem

\bibitem[\protect\citeauthoryear{Green and
  Silverman}{1994}]{green_silverman:94}
\begin{bbook}[author]
\bauthor{\bsnm{Green},~\bfnm{P.~J.}\binits{P.~J.}} \AND
  \bauthor{\bsnm{Silverman},~\bfnm{B.~W.}\binits{B.~W.}}
(\byear{1994}).
\btitle{Nonparametric regression and generalized linear models}.
\bseries{Monographs on Statistics and Applied Probability}
\bvolume{58}.
\bpublisher{Chapman \& Hall, London}
\bnote{A roughness penalty approach}.
\bdoi{10.1007/978-1-4899-4473-3}
\bmrnumber{1270012}
\end{bbook}
\endbibitem

\bibitem[\protect\citeauthoryear{Groeneboom}{2018}]{github:18}
\begin{bmisc}[author]
\bauthor{\bsnm{Groeneboom},~\bfnm{Piet}\binits{P.}}
(\byear{2018}).
\btitle{Algorithms for computing estimates in the single index model}.
\bhowpublished{\url{https://github.com/pietg/single_index}}.
\end{bmisc}
\endbibitem

\bibitem[\protect\citeauthoryear{Groeneboom and
  Jongbloed}{2014}]{piet_geurt:14}
\begin{bbook}[author]
\bauthor{\bsnm{Groeneboom},~\bfnm{Piet}\binits{P.}} \AND
  \bauthor{\bsnm{Jongbloed},~\bfnm{Geurt}\binits{G.}}
(\byear{2014}).
\btitle{Nonparametric Estimation under Shape Constraints}.
\bpublisher{Cambridge Univ. Press}, \baddress{Cambridge}.
\end{bbook}
\endbibitem

\bibitem[\protect\citeauthoryear{Han}{1987}]{han:87}
\begin{barticle}[author]
\bauthor{\bsnm{Han},~\bfnm{Aaron~K}\binits{A.~K.}}
(\byear{1987}).
\btitle{Non-parametric analysis of a generalized regression model: the maximum
  rank correlation estimator}.
\bjournal{Journal of Econometrics}
\bvolume{35}
\bpages{303--316}.
\end{barticle}
\endbibitem

\bibitem[\protect\citeauthoryear{Hristache, Juditsky and
  Spokoiny}{2001}]{hristache01}
\begin{barticle}[author]
\bauthor{\bsnm{Hristache},~\bfnm{Marian}\binits{M.}},
  \bauthor{\bsnm{Juditsky},~\bfnm{Anatoli}\binits{A.}} \AND
  \bauthor{\bsnm{Spokoiny},~\bfnm{Vladimir}\binits{V.}}
(\byear{2001}).
\btitle{Direct estimation of the index coefficient in a single-index model}.
\bjournal{Annals of Statistics}
\bpages{595--623}.
\end{barticle}
\endbibitem

\bibitem[\protect\citeauthoryear{Kuchibhotla and
  Patra}{2017}]{Kuchibhotla_patra:17}
\begin{bmisc}[author]
\bauthor{\bsnm{Kuchibhotla},~\bfnm{Arun~Kumar}\binits{A.~K.}} \AND
  \bauthor{\bsnm{Patra},~\bfnm{Rohit~Kumar}\binits{R.~K.}}
(\byear{2017}).
\btitle{Efficient Estimation in Single Index Models through Smoothing splines}.
\bhowpublished{available at \url{https://arxiv.org/abs/1612.00068}}.
\end{bmisc}
\endbibitem

\bibitem[\protect\citeauthoryear{Tanaka}{2008}]{tanaka2008}
\begin{btechreport}[author]
\bauthor{\bsnm{Tanaka},~\bfnm{Hisatoshi}\binits{H.}}
(\byear{2008}).
\btitle{Semiparametric least squares estimation of monotone single index models
  and its application to the iterative least squares estimation of binary
  choice models}
\btype{Technical Report}.
\end{btechreport}
\endbibitem

\end{thebibliography}
\end{document}